\documentclass[journal]{IEEEtran-v17}

\pdfoutput=1

\usepackage{ifthen}
\newboolean{confversion}
\setboolean{confversion}{false}

\newboolean{yurydecrepitlatex}
\ifthenelse{\boolean{confversion}}
{\setboolean{yurydecrepitlatex}{false} 
}
{
\setboolean{yurydecrepitlatex}{true} 
}

\usepackage{xspace}
\usepackage{url}
\usepackage[cmex10]{amsmath}
\usepackage{bbm}
\usepackage[dvips]{graphicx}
  \DeclareGraphicsExtensions{.eps,.pdf}
\usepackage{paralist}
\usepackage[normalem]{ulem}

\ifthenelse {\boolean{yurydecrepitlatex}}
{
}
{
  \ifCLASSOPTIONdraftcls
  \usepackage{todonotes}
  \else
  \usepackage[disable]{todonotes}
  \fi
}

\usepackage{fancyref} 
\usepackage{amsmath}
\usepackage{amssymb}
\usepackage{dsfont}
\usepackage{footnote}
\usepackage[nosort]{cite}
\usepackage{color}
\usepackage{balance}

\usepackage{bm}
\usepackage{upgreek}
\usepackage{amssymb}
\usepackage{amsfonts}
\usepackage{mathrsfs}
\usepackage{xspace}

\newcommand{\lefto}{\mathopen{}\left}
\newcommand{\safemath}[2]{\newcommand{#1}{\ensuremath{#2}\xspace}}
\safemath{\opE}{\mathbb{E}}
\newcommand{\Ex}[2]{\ensuremath{\opE_{#1}\lefto[#2\right]}} 	
\safemath{\prob}{\mathbb{P}}
\safemath{\bigO}{\mathcal{O}}
\safemath{\littleo}{\mathit{o}}

\newcommand{\tp}[1]{\ensuremath{#1^{\mathsf{T}}}} 		
\newcommand{\herm}[1]{\ensuremath{#1^{\mathsf{H}}}} 	
\newcommand{\conjugate}[1]{\ensuremath{\overline{#1}}}  
\newcommand{\inprod}[2]{\ensuremath{<\!#1,#2\!>}} 

\newtheorem{thm}{Theorem}
\newtheorem{lemma}[thm]{Lemma}
\newtheorem{dfn}{Definition}

\newcommand{\indfun}[1]{\mathbbmss{1}\{#1\}}
\newtheorem{cor}[thm]{Corollary}

\safemath{\matA}{\mathsf{A}}
\safemath{\matB}{\mathsf{B}}
\safemath{\matC}{\mathsf{C}}
\safemath{\matD}{\mathsf{D}}
\safemath{\matE}{\mathsf{E}}
\safemath{\matF}{\mathsf{F}}
\safemath{\matG}{\mathsf{G}}
\safemath{\matH}{\mathsf{H}}
\safemath{\matI}{\mathsf{I}}
\safemath{\matJ}{\mathsf{J}}
\safemath{\matK}{\mathsf{K}}
\safemath{\matL}{\mathsf{L}}
\safemath{\matM}{\mathsf{M}}
\safemath{\matN}{\mathsf{N}}
\safemath{\matO}{\mathsf{O}}
\safemath{\matP}{\mathsf{P}}
\safemath{\matQ}{\mathsf{Q}}
\safemath{\matR}{\mathsf{R}}
\safemath{\matS}{\mathsf{S}}
\safemath{\matT}{\mathsf{T}}
\safemath{\matU}{\mathsf{U}}
\safemath{\matV}{\mathsf{V}}
\safemath{\matW}{\mathsf{W}}
\safemath{\matX}{\mathsf{X}}
\safemath{\matY}{\mathsf{Y}}
\safemath{\matZ}{\mathsf{Z}}

\safemath{\randveca}{\bm{A}}
\safemath{\randvecb}{\bm{B}}
\safemath{\randvecc}{\bm{C}}
\safemath{\randvecd}{\bm{D}}
\safemath{\randvece}{\bm{E}}
\safemath{\randvecf}{\bm{F}}
\safemath{\randvecg}{\bm{G}}
\safemath{\randvech}{\bm{H}}
\safemath{\randveci}{\bm{I}}
\safemath{\randvecj}{\bm{J}}
\safemath{\randveck}{\bm{K}}
\safemath{\randvecl}{\bm{L}}
\safemath{\randvecm}{\bm{M}}
\safemath{\randvecn}{\bm{N}}
\safemath{\randveco}{\bm{O}}
\safemath{\randvecp}{\bm{P}}
\safemath{\randvecq}{\bm{Q}}
\safemath{\randvecr}{\bm{R}}
\safemath{\randvecs}{\bm{S}}
\safemath{\randvect}{\bm{T}}
\safemath{\randvecu}{\bm{U}}
\safemath{\randvecv}{\bm{V}}
\safemath{\randvecw}{\bm{W}}
\safemath{\randvecx}{\bm{X}}
\safemath{\randvecy}{\bm{Y}}
\safemath{\randvecz}{\bm{Z}}

\safemath{\randmatA}{\mathbb{A}}
\safemath{\randmatB}{\mathbb{B}}
\safemath{\randmatC}{\mathbb{C}}
\safemath{\randmatD}{\mathbb{D}}
\safemath{\randmatE}{\mathbb{E}}
\safemath{\randmatF}{\mathbb{F}}
\safemath{\randmatG}{\mathbb{G}}
\safemath{\randmatH}{\mathbb{H}}
\safemath{\randmatI}{\mathbb{I}}
\safemath{\randmatJ}{\mathbb{J}}
\safemath{\randmatK}{\mathbb{K}}
\safemath{\randmatL}{\mathbb{L}}
\safemath{\randmatM}{\mathbb{M}}
\safemath{\randmatN}{\mathbb{N}}
\safemath{\randmatO}{\mathbb{O}}
\safemath{\randmatP}{\mathbb{P}}
\safemath{\randmatQ}{\mathbb{Q}}
\safemath{\randmatR}{\mathbb{R}}
\safemath{\randmatS}{\mathbb{S}}
\safemath{\randmatT}{\mathbb{T}}
\safemath{\randmatU}{\mathbb{U}}
\safemath{\randmatV}{\mathbb{V}}
\safemath{\randmatW}{\mathbb{W}}
\safemath{\randmatX}{\mathbb{X}}
\safemath{\randmatY}{\mathbb{Y}}
\safemath{\randmatZ}{\mathbb{Z}}

\safemath{\pdff}{f}
\safemath{\pdfp}{p}
\safemath{\pdfq}{q}

\safemath{\cdfF}{F}
\safemath{\cdfP}{P}
\safemath{\cdfQ}{Q}

\safemath{\veca}{\bm{a}}
\safemath{\vecb}{\bm{b}}
\safemath{\vecc}{\bm{c}}
\safemath{\vecd}{\bm{d}}
\safemath{\vece}{\bm{e}}
\safemath{\vecf}{\bm{f}}
\safemath{\vecg}{\bm{g}}
\safemath{\vech}{\bm{h}}
\safemath{\veci}{\bm{i}}
\safemath{\vecj}{\bm{j}}
\safemath{\veck}{\bm{k}}
\safemath{\vecl}{\bm{l}}
\safemath{\vecm}{\bm{m}}
\safemath{\vecn}{\bm{n}}
\safemath{\veco}{\bm{o}}
\safemath{\vecp}{\bm{p}}
\safemath{\vecq}{\bm{q}}
\safemath{\vecr}{\bm{r}}
\safemath{\vecs}{\bm{s}}
\safemath{\vect}{\bm{t}}
\safemath{\vecu}{\bm{u}}
\safemath{\vecv}{\bm{v}}
\safemath{\vecw}{\bm{w}}
\safemath{\vecx}{\bm{x}}
\safemath{\vecy}{\bm{y}}
\safemath{\vecz}{\bm{z}}

\safemath{\matSigma}{\bm{\Sigma}}

\safemath{\setA}{\mathcal{A}}
\safemath{\setB}{\mathcal{B}}
\safemath{\setC}{\mathcal{C}}
\safemath{\setD}{\mathcal{D}}
\safemath{\setE}{\mathcal{E}}
\safemath{\setF}{\mathcal{F}}
\safemath{\setG}{\mathcal{G}}
\safemath{\setH}{\mathcal{H}}
\safemath{\setI}{\mathcal{I}}
\safemath{\setJ}{\mathcal{J}}
\safemath{\setK}{\mathcal{K}}
\safemath{\setL}{\mathcal{L}}
\safemath{\setM}{\mathcal{M}}
\safemath{\setN}{\mathcal{N}}
\safemath{\setO}{\mathcal{O}}
\safemath{\setP}{\mathcal{P}}
\safemath{\setQ}{\mathcal{Q}}
\safemath{\setR}{\mathcal{R}}
\safemath{\setS}{\mathcal{S}}
\safemath{\setT}{\mathcal{T}}
\safemath{\setU}{\mathcal{U}}
\safemath{\setV}{\mathcal{V}}
\safemath{\setW}{\mathcal{W}}
\safemath{\setX}{\mathcal{X}}
\safemath{\setY}{\mathcal{Y}}
\safemath{\setZ}{\mathcal{Z}}
\safemath{\emptySet}{\varnothing}

\safemath{\veczero}{\mathbf{0}} 

\safemath{\diag}{\mathrm{diag}}

\safemath{\jpg}{\mathcal{CN}}			
\safemath{\complexset}{\mathbb{C}}

\usepackage{bm}
\usepackage{upgreek}
\usepackage{amssymb}
\usepackage{amsfonts}
\usepackage{mathrsfs}
\usepackage{xspace}

\safemath{\mi}{I}
\safemath{\difent}{\mathrm{h}}		
\safemath{\NonnegReal}{\mathbb{R}^{+}}

\safemath{\indist}{\cdfP} 
\safemath{\outdist}{\cdfQ} 
\safemath{\inpdf}{\pdfp} 
\safemath{\outpdf}{\pdfq} 
\safemath{\testdist}{\cdfP} 

\safemath{\inset}{\setF} 

\safemath{\encoder}{f} 
\safemath{\decoder}{g} 
\safemath{\msg}{J} 
\safemath{\csir}{\mathrm{r}}
\safemath{\csit}{\mathrm{t}}
\safemath{\csi}{\mathrm{csi}}
\safemath{\csirt}{\mathrm{rt}}
\safemath{\Rcsirt}{R_{\csirt}} 
\safemath{\Rcsir}{R_{\csir}}
\safemath{\Rcsit}{R_{\csit}}

\safemath{\Rnocsit}{R_{\mathrm{no}}} 
\safemath{\Rnocsi}{R_{\mathrm{no}}} 
\safemath{\Rgeneral}{R_{\mathrm{no,rt}}} 
\safemath{\equalR}{R_{\mathrm{e}}}

\safemath{\Vnocsit}{V_{\error}^{\mathrm{no}}}
\safemath{\Vcsit}{V_{\error}^{\mathrm{rt}}}

\safemath{\cadist}{F_C} 

\safemath{\deF}{d_0} 

\safemath{\bl}{n} 
\safemath{\error}{\epsilon} 
\safemath{\NumCode}{M}
\safemath{\RXant}{r}
\safemath{\rxant}{r}
\safemath{\snr}{\rho}

\safemath{\BoundFU}{k_{\delta}}

\safemath{\argpn}{\xi} 

\safemath{\angletest}{Z} 

\safemath{\errorach}{P_\mathrm{e}} 

\newcommand{\given}{\,\vert\,}				
\safemath{\define}{\triangleq}			
\safemath{\fnorm}{\mathrm{F}}

\safemath{\altbl}{\tilde{\bl}}
\safemath{\constL}{k_{\mathrm{L}}}
\safemath{\constU}{k_{\mathrm{U}}}
\safemath{\funcL}{\tilde{q}}
\safemath{\funcU}{q}
\safemath{\ConstThm}{k_0}
\safemath{\randrevec}{\randvecy}
\safemath{\revec}{\vecy}
\safemath{\trcwd}{\vecx}
\safemath{\randtrcwd}{\randvecx}
\safemath{\randnoisevec}{\randvecw}
\safemath{\transmitcwd}{\vecx_1} 

\safemath{\pickcwdnoch}{\vecx_0} 
\safemath{\pickcwd}{\vecx_0} 

\safemath{\inseqrand}{\randvecx}
\safemath{\inseq}{\vecx}

\safemath{\outseq}{\matY}
\safemath{\outseqrand}{\randmatY}

\safemath{\altT}{\widetilde{T}}
\safemath{\altU}{\widetilde{U}}
\safemath{\altmean}{\tilde{\mu}}
\safemath{\altvar}{\tilde{\sigma}}
\safemath{\altf}{\funcL}
\safemath{\altg}{\tilde{g}}
\safemath{\altgamma}{\tilde{\gamma}}
\safemath{\altdelta}{\tilde{\delta}}
\safemath{\altk}{\tilde{k}}
\safemath{\constant}{\tilde{k}}

\def\PP{\prob}

\begin{document}
\IEEEoverridecommandlockouts

\title{Quasi-Static SIMO Fading Channels\\ at Finite Blocklength}

\author{\IEEEauthorblockN{Wei Yang$^1$, Giuseppe Durisi$^1$,  Tobias Koch$^2$, and Yury Polyanskiy$^3$}
\thanks{This work was supported by the National Science Foundation under Grant CCF-1253205, by a Marie Curie FP7 Integration Grant within the 7th European Union Framework Programme under Grant 333680, by the Spanish government (TEC2009-14504-C02-01, CSD2008-00010, and TEC2012-38800-C03-01), and by an Ericsson's Research Foundation grant.}
\\
\IEEEauthorblockA{
$^1$Chalmers University of Technology, 41296 Gothenburg, Sweden\\
$^2$Universidad Carlos III de Madrid, 28911 Legan\'{e}s, Spain\\
$^3$Massachusetts Institute of Technology, Cambridge, MA, 02139 USA
}}

\maketitle

\begin{abstract}
We investigate the maximal achievable rate for a given blocklength and error probability over quasi-static single-input multiple-output (SIMO) fading channels.
Under mild conditions on the channel gains, it is shown that the channel dispersion is zero regardless of whether the fading realizations are available at the transmitter and/or the receiver.
The result follows from computationally and analytically tractable converse and achievability bounds.
Through numerical evaluation, we verify that, in some scenarios, zero dispersion indeed entails fast convergence to outage capacity as the blocklength increases.
In the example of a particular $1\times 2$ SIMO Rician channel, the blocklength required to achieve $90\%$ of capacity is about an order of magnitude smaller compared to the blocklength required for an AWGN channel with the same capacity.

\end{abstract}

\section{Introduction}
We study the maximal achievable rate $R^{*}(\bl, \error)$ for a given blocklength $\bl$ and block error probability $\error$ over a \emph{quasi-static} single-input multiple-output (SIMO) fading channel, i.e., a random channel that remains constant during the transmission of each codeword, subject to a per-codeword power constraint.
We consider two scenarios:
\ifthenelse{\boolean{confversion}}
{
\begin{inparaenum}[i)]
\item perfect channel-state information (CSI) is available at both the transmitter and the receiver;\footnote{Hereafter, we write CSIT and CSIR to denote the availability of perfect CSI at the transmitter and at the receiver, respectively.
The acronym CSIRT will be used to denote the availability of both CSIR and CSIT.}
\item neither the transmitter nor the receiver have \emph{a priori} CSI.
\end{inparaenum} }
{
\begin{enumerate}[i)]
\item perfect channel-state information (CSI) is available at both the transmitter and the receiver;\footnote{Hereafter, we write CSIT and CSIR to denote the availability of perfect CSI at the transmitter and at the receiver, respectively.
The acronym CSIRT will be used to denote the availability of both CSIR and CSIT.}
\item neither the transmitter nor the receiver have \emph{a priori} CSI.
\end{enumerate}
}

For quasi-static fading channels, the Shannon capacity, which is the limit of $R^{*}(\bl, \error)$ for $\bl\to\infty$ and $\error\to0$, is zero for many fading distributions of practical interest (e.g., Rayleigh, Rician, and Nakagami fading).
In this case, the $\error$-capacity~\cite{verdu94-07a} (also known as \emph{outage} capacity), which is obtained by letting $\bl\to\infty$ in $R^{*}(\bl, \error)$ for a fixed $\error >0$, is a more appropriate performance metric.
The $\error$-capacity of quasi-static SIMO fading channels does not depend on whether CSI is available at the  receiver~\cite[p.~2632]{biglieri98-10a}.
 In fact, since the channel stays constant during the transmission of a codeword, it can be accurately estimated at the receiver through the transmission of known training sequences with no rate penalty as $\bl\to\infty$.
 Furthermore, in the limit $n\to\infty$ the per-codeword power constraint renders CSIT
ineffectual~\cite[Prop.~3]{caire99-05}, in contrast to the situation where a long-term power constraint is imposed~\cite{caire99-05,goldsmith97-11a}.

Building upon classical asymptotic results of Dobrushin and Strassen, it was recently shown by Polyanskiy, Poor, and Verd\'u~\cite{polyanskiy10-05} that for various channels with positive Shannon capacity $C$, 
the maximal achievable rate can be tightly approximated by
\vspace{-1mm}
\begin{IEEEeqnarray}{rCl}
    \label{eq:approx_R_introduction}
    R^{*}(\bl,\error)  = C -\sqrt{\frac{V}{\bl}}Q^{-1}(\error) + \bigO\lefto(\frac{\log \bl }{\bl}\right).
\end{IEEEeqnarray}
Here, $Q^{-1}(\cdot)$ denotes the inverse of the Gaussian $Q$-function and $V$ is the \emph{channel dispersion}~\cite[Def.~1]{polyanskiy10-05}.
 The approximation~(\ref{eq:approx_R_introduction}) implies that to sustain the desired error probability~$\error$ at a finite blocklength $\bl$, one pays a penalty on the rate (compared to the channel capacity) that is proportional to~$1/\sqrt{\bl}$.
\ifthenelse{\boolean{confversion}}
{}
{For the CSIR case, the dispersion of single-input single-output AWGN channels with stationary fading was derived in \cite{polyanskiy11-isit}, and generalized to block-memoryless fading channels in \cite{yang12-09}.}

\paragraph*{Contributions}
We provide achievability and converse bounds on $R^\ast(\bl,\error)$ for quasi-static SIMO fading channels.
The asymptotic analysis of these bounds shows that under mild technical conditions on the distribution of the fading gains,
\ifthenelse{\boolean{confversion}}
{    \begin{IEEEeqnarray}{rCl}
    \label{eq:approx_R_introduction-quasi-static}
    R^{*}(\bl,\error)  = C_{\epsilon} + \bigO\lefto({\log(\bl)/ \bl}\right).
    \end{IEEEeqnarray}}
{    \begin{IEEEeqnarray}{rCl}
    \label{eq:approx_R_introduction-quasi-static}
    R^{*}(\bl,\error)  = C_{\epsilon} + \bigO\lefto({\log\bl\over \bl}\right).
    \end{IEEEeqnarray}}
This result implies that for the quasi-static fading case, the $1/\sqrt{\bl}$ rate penalty is absent.
In other words, the \error-dispersion (see~\cite[Def.~2]{polyanskiy10-05} or \eqref{eq:def-dispersion-epsilon} below) of quasi-static fading channels is zero.
This result turns out to hold regardless of whether CSI is available at the transmitter and/or the receiver.

Numerical evidence suggests that, in some scenarios, the absence of the $1/\sqrt{n}$ term in~\eqref{eq:approx_R_introduction-quasi-static} implies fast convergence to $C_{\error}$ as $n$ increases.
For example, for a $1\times 2$ SIMO Rician-fading channel with $C_{\epsilon}=1$ bit$/$channel use and $\epsilon=10^{-3}$, the blocklength required to achieve $90\%$ of $C_{\epsilon}$ is between $120$ and $320$, which is about an order of magnitude smaller compared to the blocklength required for an AWGN channel with the same capacity.
%
In general, to estimate $R^{*}(\bl,\error)$ accurately for moderate \bl, an asymptotic characterization more precise than~\eqref{eq:approx_R_introduction-quasi-static} is required.

Our converse bound on $R^{*}(\bl,\error)$ is based on the meta-converse theorem~\cite[Thm.~26]{polyanskiy10-05}.
Application of standard achievability bounds for the case of no CSI encounters formidable technical and numerical difficulties. To circumvent them, we apply the $\kappa\beta$ bound~\cite[Thm.~25]{polyanskiy10-05} to a stochastically degraded channel, whose choice is motivated by geometric considerations.
%
The main tool used to establish~\eqref{eq:approx_R_introduction-quasi-static} is a Cramer-Esseen-type central-limit theorem~\cite[Thm.~VI.1]{petrov75}.

\paragraph*{Notation} 
\label{sec:notation}
Upper case letters denote scalar random variables and lower case letters denote their realizations.
We use boldface upper case letters to denote random vectors, e.g., $\randvecx$, and boldface lower case letters for their realizations, e.g., $\vecx$.
Upper case letters of two special fonts are used to denote deterministic matrices (e.g., $\matY$) and random matrices (e.g., $\randmatY$).
%
\ifthenelse{\boolean{confversion}}
{
The superscripts~$\tp{}$ and $\herm{}$ stand for transposition and Hermitian transposition, respectively.
}
{
The element-wise complex conjugate of the vector $\vecx$ is denoted by $\conjugate{\vecx}$.
The superscripts~$\tp{}$ and $\herm{}$ stand for transposition and Hermitian transposition, respectively.
The standard (Hermitian) inner product of two vectors $\vecx=\tp{[x_1\, \cdots\, x_\bl]}$ and $\vecy=\tp{[y_1\, \cdots\, y_\bl]}$ is
\begin{IEEEeqnarray}{rCl}
\langle\vecx,\vecy\rangle \define \sum\limits_{i=1}^{\bl} x_i \conjugate{y_i}.
\end{IEEEeqnarray}
The Euclidean norm is denoted by $\|\vecx\|^2 \define \langle\vecx,\vecx\rangle$. }
Furthermore, $\jpg(\mathbf{0}, \matA)$ stands for the distribution of a circularly-symmetric complex
Gaussian random vector with covariance matrix~$\matA$.
\ifthenelse{\boolean{confversion}}
{}
{Given two distributions $\indist$ and $\outdist$ on a common measurable space $\setW$, we define a randomized test between $\indist$ and $\outdist$ as a random transformation $\testdist_{Z\given W}: \setW\mapsto\{0,1\}$ where $0$ indicates that the test chooses $\outdist$. We shall need the following performance metric for the test between~$\indist$ and~$\outdist$:
\begin{IEEEeqnarray}{rCl}
\label{eq:def-beta}
\beta_\alpha(\indist,\outdist) \define \min\int \testdist_{Z\given W}(1\given w)  \outdist(d w)
\end{IEEEeqnarray}
where the minimum is over all probability distributions $\testdist_{Z\given W}$
satisfying
\begin{IEEEeqnarray}{rCl}
\int \testdist_{Z\given W} (1\given w) \indist(dw)\geq \alpha.
\end{IEEEeqnarray}
We refer to a test achieving (\ref{eq:def-beta}) as an optimal test.}
The indicator function is denoted by $\indfun{\cdot}$.
Finally, $\log(\cdot)$ indicates the natural logarithm, and $\mathrm{Beta}(\cdot,\cdot)$ denotes the Beta distribution~\cite[Ch.~25]{johnson95-2}.

\section{Channel Model and Fundamental Limits}
\label{sec:system_model}
We consider a quasi-static SIMO channel with $\rxant$ receive antennas. The channel input-output relation is given by
%
%
\begin{IEEEeqnarray}{rCl}
\label{eq:channel_io_new}
\randmatY &=&  \vecx \tp{\randvech} + \randmatW\\
&=&
\begin{pmatrix}
  x_1 H_1 + W_{11} & \cdots & x_1 H_r + W_{1r} \\
  \vdots &  & \vdots \\
  x_{\bl} H_1 + W_{n1} & \cdots & x_{\bl} H_r + W_{nr}
\end{pmatrix}
.
\end{IEEEeqnarray}
The vector $\randvech=\tp{[H_1\,\cdots\,H_\rxant]}$ contains the complex fading coefficients, which are random but remain constant for all~$\bl$ channel uses;  $\{W_{lm}\}$ are independent and identically distributed (i.i.d.) $\jpg(0,1)$ random variables; $\vecx=\tp{[x_1\,\cdots\,x_\bl]}$ contains the transmitted symbols.

We consider both the case when the transmitter and the receiver do not know the realizations of~$\randvech$ (no CSI) and the case where
the realizations of $\randvech$ are available to both the transmitter and the receiver (CSIRT).
Next, we introduce the notion of a channel code for these two settings.

\begin{dfn} An $(\bl, \NumCode, \error)_{\text{no-CSI}}$ code  consists of:
\begin{enumerate}[i)]
\item an encoder $\encoder$: $\{1,\ldots,\NumCode\} \mapsto \complexset^\bl$ that maps the message $\msg \in \{1,\ldots,\NumCode\}$ to a codeword $\inseq \in \{\vecc_1,\ldots, \vecc_{\NumCode}\}$. The codewords satisfy the power constraint
\begin{IEEEeqnarray}{rCl}
\label{eq:peak-power-constraint}
\|\vecc_i\|^2\leq \bl \snr,\quad i=1,\ldots,\NumCode.
\end{IEEEeqnarray}
We assume that $\msg$ is equiprobable on $\{1,\ldots,\NumCode\}$.
\item A decoder $\decoder$: $\complexset^{\bl\times \rxant } \mapsto\{1,\ldots,\NumCode\}$ satisfying
\ifthenelse{\boolean{confversion}}
{$\prob[\decoder(\outseqrand) \neq \msg ] \leq \error$, }{
\begin{equation}
\prob[\decoder(\outseqrand) \neq \msg ] \leq \error
\end{equation}}
where $\outseqrand$ is the channel output induced by the transmitted codeword according to~\eqref{eq:channel_io_new}.
\end{enumerate}
The maximal achievable rate for the no-CSI case is defined as
%
%
\begin{equation}
\label{eq:def-r-nocsit}
\Rnocsit^{\ast}(\bl,\error) \define \sup\lefto\{\frac{\log\NumCode}{\bl}\!: \exists(\bl
,\NumCode, \error)_{\text{no-CSI}}\text{ code}\right\}.
\end{equation}
%
%
 \end{dfn}

 \begin{dfn} An $(\bl, \NumCode, \error)_{\text{CSIRT}}$ code consists of:
\begin{enumerate}[i)]
\item an encoder $\encoder$: $ \{1,\ldots,\NumCode\}\times \complexset^{\rxant} \mapsto \complexset^\bl$ that maps the message $\msg \in \{1,\ldots,\NumCode\}$ and the channel $\randvech$ to a codeword
$\inseq \in \{\vecc_1{(\randvech)},\ldots,\vecc_\NumCode{(\randvech)} \}$.
 The codewords satisfy the power constraint
 \begin{IEEEeqnarray}{c}
 \label{eq:power-constraint-csirt}
   \|\vecc_i(\vech)\|^2\leq \bl\snr, \quad \forall i=1,\dots \NumCode,\quad \forall \vech \in \complexset^{r}.
 \end{IEEEeqnarray}
  We assume that $\msg$ is equiprobable on $\{1,\ldots,\NumCode\}$.
\item A decoder $\decoder$: $\complexset^{\bl\times \rxant } \times \complexset^\rxant \mapsto\{1,\ldots,\NumCode\}$ satisfying
\ifthenelse{\boolean{confversion}}
{$\prob[\decoder(\outseqrand,\randvech) \neq \msg ] \leq \error$.}{ \begin{equation}
\prob[\decoder(\outseqrand,\randvech) \neq \msg ] \leq \error.
\end{equation}}
\end{enumerate}
The maximal achievable rate for the CSIRT case is defined as
%
%
\begin{equation}
\Rcsirt^{\ast}(\bl,\error) \define\sup\lefto\{\frac{\log\NumCode}{\bl}\!: \exists(\bl
,\NumCode, \error)_{\text{CSIRT}}\text{ code}\right\}.
\end{equation}
%
%
 \end{dfn}
 \setlength{\baselineskip}{\normalbaselineskip}
It follows that\ifthenelse{\boolean{confversion}}
{ $\Rnocsit^{\ast}(\bl,\error) \leq \Rcsirt^{\ast}(\bl,\error)$.}
{
\begin{IEEEeqnarray}{rCl}
\label{eq:relation-csit-nocsit}
\Rnocsit^{\ast}(\bl,\error) \leq \Rcsirt^{\ast}(\bl,\error).
\end{IEEEeqnarray}
}

 Let $G \define \|\randvech\|^2$, and define
\begin{equation}
\label{eq:def-cadist}
\cadist(\argpn) \define \prob\left[\log(1+\snr G) \leq \argpn \right].
\end{equation}
For every $\error>0$, the $\error$-capacity $C_\error$ of the channel~\eqref{eq:channel_io_new} is~\cite[Thm.~6]{verdu94-07a}
\begin{equation}
\label{eq:epsilon-capacity}
C_\error =\!\! \lim\limits_{\bl\to\infty} \Rnocsit^{*}(\bl,\error)  =\!\! \lim\limits_{\bl\to\infty} \Rcsirt^{*}(\bl,\error) = \sup\left\{\argpn: \cadist(\argpn) \leq \error \right\}.\IEEEeqnarraynumspace
\end{equation}

\section{Main Results}
In Section~\ref{sec:converse-bound}, we present a converse (upper) bound on $\Rcsirt^\ast(\bl,\error)$ and in Section~\ref{sec:ach-bound} we present an achievability (lower) bound on $\Rnocsit^\ast(\bl,\error)$. We show in~Section~\ref{sec:asymptotic-analysis} that the two bounds match asymptotically up to a $\bigO(\log(\bl)/\bl)$ term, which allows us to establish (\ref{eq:approx_R_introduction-quasi-static}).

\subsection{Converse Bound}
\label{sec:converse-bound}
\begin{thm}
\label{thm:converse}
Let
\ifthenelse{\boolean{confversion}}{
\begin{IEEEeqnarray}{rCl}
\label{eq:info_density_simo_alt}
  L_\bl &\define& \bl \log(1+\snr G)  + \sum\limits_{i=1}^{\bl}\!\!
  \left(1-\bigl| \sqrt{\snr G}Z_i-\!  \sqrt{1+\snr G} \bigr|^2\right)\!\!\IEEEeqnarraynumspace\\
  S_\bl &\define& \bl \log(1+\snr G)  +   \sum\limits_{i=1}^{\bl}\left(1-\frac{\big|\sqrt{\snr G}Z_i-1\big|^2}{1+\snr G}\right)\label{eq:info_density_simo}
\end{IEEEeqnarray}}
{\begin{equation}
\label{eq:info_density_simo_alt}
  L_\bl \define \bl \log(1+\snr G)  + \sum\limits_{i=1}^{\bl}
  \left(1-\bigl| \sqrt{\snr G}Z_i-  \sqrt{1+\snr G} \bigr|^2\right)
\end{equation}
and
\begin{IEEEeqnarray}{rCl}
S_\bl &\define& \bl \log(1+\snr G)  +   \sum\limits_{i=1}^{\bl}\left(1-\frac{\big|\sqrt{\snr G}Z_i-1\big|^2}{1+\snr G}\right)\label{eq:info_density_simo}
\end{IEEEeqnarray}
}
with $G = \|\randvech\|^2$ and $\{Z_i\}_{i=1}^{\bl}$ i.i.d. $\jpg(0,1)$-distributed.
For  every $\bl$ and every $0<\error<1$, the maximal achievable rate on the quasi-static SIMO fading channel~\eqref{eq:channel_io_new} with CSIRT is upper-bounded by
%
%
\begin{IEEEeqnarray}{rCl}
\label{eq:thm-converse-rcsit}
\Rcsirt^{\ast}(\bl-1,\error) \leq \frac{1}{n-1}\log \frac{1}{\prob[L_\bl \geq \bl \gamma_{\bl}]}
\end{IEEEeqnarray}
where $\gamma_{\bl}$ is the solution of
\ifthenelse{\boolean{confversion}}{$\prob[ S_\bl \leq  \bl \gamma_{\bl}] =\error.$}{
\begin{IEEEeqnarray}{rCl}
 \label{eq:thm-converse-def-gamma-n}
 \prob[ S_\bl \leq  \bl \gamma_{\bl}] =\error.
 \end{IEEEeqnarray}}

\end{thm}
\ifthenelse{\boolean{confversion}}
{
\begin{IEEEproof}
See Appendix \ref{sec:app-proof-converse}.
\end{IEEEproof}
}
{
\begin{IEEEproof}
See Appendix \ref{sec:app-proof-converse}.
\end{IEEEproof}
}
%

\subsection{Achievability Bound}
\label{sec:ach-bound}
%
%
Let $Z(\matY):\,\complexset^{\bl\times\rxant} \mapsto \{0 , 1\}$ be a test between $\indist_{\randmatY | \randvecx = \vecx}$ and an arbitrary distribution $\outdist_{\randmatY}$, where $Z = 0$ indicates that the test chooses $\outdist_{\randmatY}$.
Let $\inset \subset \complexset^{\bl}$ be a set of permissible channel inputs as specified by~\eqref{eq:peak-power-constraint}.
We define the following measure of performance $\tilde{\kappa}_{\tau}(\inset, \outdist_{\randmatY})$ for the composite hypothesis test between $\outdist_{\randmatY}$ and the collection $\{\indist_{\randmatY|\randvecx = \vecx}\}_{\vecx \in \inset}$:
    \begin{IEEEeqnarray}{rCl}
     \tilde{\kappa}_{\tau}(\inset, \outdist_{\randmatY}) &\define&
    \inf \outdist_{\randmatY}\left[ Z(\randmatY)=1\right]    \label{eq:def_kappa_tilde}
\end{IEEEeqnarray}
where the infimum is over all \emph{deterministic} tests $Z(\cdot)$~satisfying:
\begin{enumerate}[i)]
\item $\indist_{\randmatY | \randvecx = \vecx}\left[ Z(\randmatY) = 1\right] \geq \tau,\,\forall \vecx \in \inset$, and\label{item:constraint1}
\item $Z(\matY) = Z(\widetilde{\matY})$ whenever the columns of $\matY$ and $\widetilde{\matY}$ span the same subspace in $\complexset^{\bl}$\label{item:constraint2}.
\end{enumerate}
Note that, $\tilde{\kappa}_{\tau}(\inset, \outdist_{\randmatY})$  in (\ref{eq:def_kappa_tilde}) coincides with $\kappa_{\tau}(\inset ,\outdist_{\randmatY})$ defined in \cite[eq.~(107)]{polyanskiy10-05} if the additional constraint \ref{item:constraint2}) is dropped and if the infimum in (\ref{eq:def_kappa_tilde}) is taken over randomized tests.
 Hence, \ifthenelse{\boolean{confversion}}{$\kappa_\tau(\inset ,\outdist_{\randmatY}) \leq \tilde{\kappa}_{\tau}(\inset ,\outdist_{\randmatY})$.}
{\begin{equation}
\label{eq:upper-bound-kappa-tau}
\kappa_\tau(\inset ,\outdist_{\randmatY}) \leq \tilde{\kappa}_{\tau}(\inset ,\outdist_{\randmatY}). 
\end{equation}
where the RHS of (\ref{eq:upper-bound-kappa-tau}) is achieved by the trivial test that sets $Z=1$ with probability $\tau$ independent of $\randmatY$.
}

To state our lower bound on $\Rnocsit^\ast(\bl,\error)$, we will need the following definition.
%
\begin{dfn}
Let $\veca$ be a nonzero vector and let $\setB$ be an $l$-dimensional ($l<\bl$) subspace in $\mathbb{C}^{\bl}$. The angle $\theta(\veca,\setB)\in[0,\pi/2]$ between $\veca$ and $\setB$ is defined by
\ifthenelse{\boolean{confversion}}
{
\begin{IEEEeqnarray}{rCl}
\label{eq:definition-angle-vec-subspace}
\cos\theta(\veca,\setB) = \max\limits_{\vecb \in \setB,\, \|\vecb\|=1}|\herm{\veca}\vecb|/\|\veca\|.
\end{IEEEeqnarray}
\end{dfn}
}
{
\begin{IEEEeqnarray}{rCl}
\label{eq:definition-angle-vec-subspace}
\cos\theta(\veca,\setB) = \max\limits_{\vecb \in \setB,\, \|\vecb\|=1}\frac{|\langle \veca, \vecb\rangle|}{\|\veca\|}.
\end{IEEEeqnarray}
\end{dfn}}
With a slight abuse of notation, for a matrix $\matB \in \mathbb{C}^{\bl\times l}$ we use~$\theta(\veca,\matB)$ to indicate the angle between $\veca$ and the subspace~$\setB$ spanned by the columns of~$\matB$. In particular, if the columns of $\matB$ are an orthonormal basis for $\setB$, then
\ifthenelse{\boolean{confversion}}
{
\begin{equation}
\label{eq:ortho-cos-theta}
\cos\theta(\veca, \matB)  = \| \herm{\veca}\matB\|/\|\veca\|.
\end{equation}
}
{
\begin{equation}
\label{eq:ortho-cos-theta}
\cos\theta(\veca, \matB)  = \frac{\| \herm{\veca}\matB\|}{\|\veca\|}.
\end{equation}
}


%

\begin{thm}
\label{thm:kappa-beta-achievability}
Let $\inset\subset \complexset^{\bl}$ be a measurable set of  channel inputs satisfying~\eqref{eq:peak-power-constraint}.
For every $0<\error<1$, every \mbox{$0<\tau<\error$}, and every probability distribution $\outdist_{\randmatY}$, there exists an $(\bl, \NumCode, \error)_{\text{no-CSI}}$ code satisfying
\begin{IEEEeqnarray}{rCl}
\label{eq:lb-numcode-simo}
\NumCode &\geq&
\frac{\tilde{\kappa}_{\tau}(\inset, \outdist_{\randmatY})}{\sup_{\vecx \in \inset}\outdist_{\randmatY}[Z_{\vecx}(\randmatY)=1]}
\end{IEEEeqnarray}
where
\begin{IEEEeqnarray}{rCl}
\label{eq:threshold_tester_Z}
\angletest_{\vecx}(\matY) &=& \indfun{\cos^2\!\theta(\vecx, \matY)\geq 1- \gamma_\bl(\vecx)}
\end{IEEEeqnarray}
with $\gamma_\bl(\vecx)\in[0,1]$ chosen so that
\begin{IEEEeqnarray}{rCl}
\label{eq:definition_gamma_x}
 \indist_{\randmatY | \randvecx = \vecx}[\angletest_{\vecx}(\randmatY)=1] \geq 1-\error+\tau.
\end{IEEEeqnarray}
\end{thm}
\ifthenelse {\boolean{confversion}}
{
\begin{IEEEproof}
The bound~(\ref{eq:lb-numcode-simo}) follows by applying the $\kappa\beta$ bound~\cite[Thm.~25]{polyanskiy10-05} to a stochastically degraded version of~(\ref{eq:channel_io_new}), whose output is the subspace spanned by the columns of~$\randmatY$.
\end{IEEEproof}
}
{
\begin{IEEEproof}
The lower bound (\ref{eq:lb-numcode-simo}) follows by applying the $\kappa\beta$ bound~\cite[Thm.~25]{polyanskiy10-05} to a stochastically degraded version of (\ref{eq:channel_io_new}), whose output is the subspace spanned by the columns of~$\randmatY$.
\end{IEEEproof}
}

The geometric intuition behind the choice of the test (\ref{eq:threshold_tester_Z}) is that $\inseq$ in (\ref{eq:channel_io_new}) belongs to the subspace spanned by the columns of $\randmatY$ if the additive noise $\randmatW$ is neglected.



In Corollary~\ref{cor:actual_achievability_bound} below, we present a further lower bound on~$\NumCode$ that is obtained from Theorem~\ref{thm:kappa-beta-achievability} by choosing 
\begin{IEEEeqnarray}{rCl}\label{eq:Gaussian_output_distribution}
\outdist_{\randmatY} &=& \prod\limits_{i=1}^{\bl}\jpg(\mathbf{0},\matI_\rxant)
\end{IEEEeqnarray}
and by requiring that the codewords belong to the set
\begin{IEEEeqnarray}{c}\label{eq:equal_power_constraint}
\inset_\bl \define \left\{\inseq\in \complexset^{\bl}: \|\inseq\|^2 = \bl\snr \right\}.
\end{IEEEeqnarray}
The resulting bound allows for numerical evaluation.
\begin{cor}\label{cor:actual_achievability_bound}
 For every $0\!<\!\error\!<\!1$ and every $0\!<\!\tau\!<\!\error$ there exists an $(\bl, \NumCode, \error)_{\text{no-CSI}}$ code with codewords in the set~$\inset_\bl$ satisfying
 \begin{IEEEeqnarray}{rCl}
 \label{eq:lb-numcode-simo_final}
 \NumCode &\geq& \frac{\tau}{F(\gamma_{\bl};\bl-\rxant,\rxant) }
 \end{IEEEeqnarray}
 where $F(\cdot;\bl-\rxant,\rxant)$
 is the cumulative distribution function (cdf) of a $\mathrm{Beta}(\bl-\rxant,\rxant)$-distributed random variable and \mbox{$\gamma_{\bl}\in[0,1]$} is chosen so that
 %
 %
 \begin{IEEEeqnarray}{rCl}
  \indist_{\randmatY | \randvecx = \vecx_0}[\angletest_{\vecx_0}(\randmatY)=1] \geq 1-\error+\tau
 \end{IEEEeqnarray}
with
\begin{IEEEeqnarray}{c}\label{eq:vec0}
\pickcwd \define \tp{\big[\sqrt{\snr}\,\sqrt{\snr} \,\cdots \,\sqrt{\snr}\big]}.
\end{IEEEeqnarray}
\end{cor}
\ifthenelse{\boolean{confversion}}
{
\begin{IEEEproof}
See Appendix~\ref{app:corollary}.
\end{IEEEproof}
}{
\begin{IEEEproof}
See Appendix~\ref{app:corollary}.
\end{IEEEproof}
}
 %


\subsection{Asymptotic Analysis}
\label{sec:asymptotic-analysis}
\setlength{\baselineskip}{1.18em}
{Following~\cite[Def. 2]{polyanskiy10-05}, we define the $\error$-dispersion of the
channel~\eqref{eq:channel_io_new} via $\Rnocsit^{\ast}(\bl,\error)$ (resp.
$\Rcsirt^{\ast}(\bl,\error)$) as
\begin{IEEEeqnarray}{rCl}
\label{eq:def-dispersion-epsilon}
V_\error^{\mathrm{no}} &\define& \limsup\limits_{\bl\to\infty} \bl\left( \frac{C_\error -
\Rnocsit^{\ast}(n,\epsilon)}{Q^{-1}(\error)}\right)^2,\, \error\in(0,1)\backslash \Big\{\frac{1}{2}\Big\} \\
V_\error^{\mathrm{rt}} &\define& \limsup\limits_{\bl\to\infty} \bl\left( \frac{C_\error -
\Rcsirt^{\ast}(n,\epsilon)}{Q^{-1}(\error)}\right)^2,\, \error\in(0,1)\backslash \Big\{\frac{1}{2}\Big\} .\IEEEeqnarraynumspace
\end{IEEEeqnarray}
The rationale behind the definition of the channel dispersion is that---for ergodic channels---the
probability of error $\epsilon$ and the optimal rate $R^*(n, \epsilon)$ roughly satisfy
\ifthenelse{\boolean{confversion}}
{
\begin{equation}\label{eq:int1}
	 \epsilon \approx \PP\lefto[C + \sqrt{V/\bl}\, Z \le R^*(n, \epsilon)\right]
\end{equation}}
{
\begin{equation}\label{eq:int1}
	 \epsilon \approx \PP\lefto[C + \sqrt{\frac{V}{\bl}}\, Z \le R^*(n, \epsilon)\right]
\end{equation}}
where $C$ and $V$ are the channel capacity and dispersion, respectively, and $Z$ is a zero-mean unit-variance real Gaussian random variable.
The quasi-static fading channel is conditionally ergodic given $\bf H$, which suggests that
\ifthenelse{\boolean{confversion}}{
\begin{equation}\label{eq:int2}
	 \epsilon \approx \PP\lefto[C(\mathbf{H}) + \sqrt{V(\mathbf{H})/ n}\, Z \le R^*(n, \epsilon)\right]
\end{equation}}{
\begin{IEEEeqnarray}{rCl}
\label{eq:int2}
	 \epsilon &\approx& \PP\lefto[C(\mathbf{H}) + \sqrt{\frac{V(\mathbf{H})}{ n}}\, Z \le R^*(n, \epsilon)\right]
\end{IEEEeqnarray}}
where $C(\randvech)$ and $V(\randvech)$ are the capacity and the dispersion of the
conditional channels.
Assume that $Z$ is independent of $\randvech$. Then, given $\mathbf{H}=\vech$,   the probability
$\PP[Z \le (R^*(n, \epsilon)-C(\mathbf{\vech}))/\sqrt{V(\vech)/n}]$ is close to one in the ``outage'' case $C(\mathbf{\vech})< R^*(n, \epsilon)$, and  close to zero otherwise. Hence, we
expect that~\eqref{eq:int2} be well-approximated by
\begin{equation}\label{eq:int3}
\error \approx \PP\lefto[C(\mathbf{H}) \le R^*(n, \epsilon)\right].
\end{equation}
This observation is formalized in the following lemma.}
\ifthenelse{\boolean{confversion}}{
}
{
\begin{figure*}[t]
	\centering
		\includegraphics[scale=1.0]{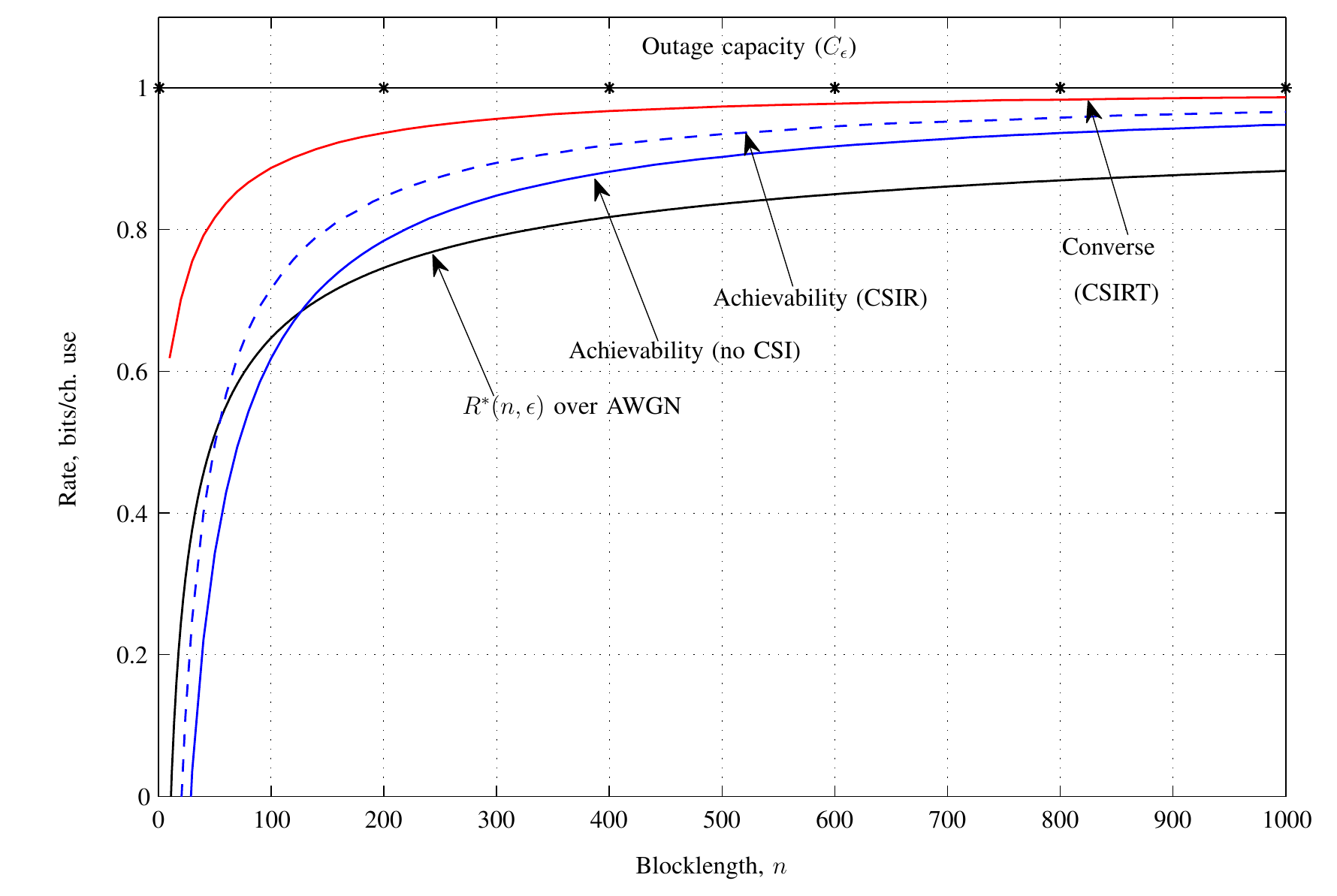}
\caption{Bounds for the quasi-static SIMO Rician-fading channel with $K$-factor equal to $20$ dB, two receive antennas, $\text{SNR}=-1.55 $ dB, and $\epsilon=10^{-3}$.\label{fig:gamma_distribution}}
\end{figure*}
}
\begin{lemma}
\label{lem:expectation-phi}
Let $A$ be a random variable with zero mean, unit variance, and finite third moment.
Let $B$ be independent of $A$ with twice continuously differentiable probability density function (pdf) $\pdff_B$.
Then, there exists $k_1<\infty$ such that
\begin{IEEEeqnarray}{rCl}
\lim \limits_{ \bl\to\infty } n^{3/2} \left|\prob[A\leq \sqrt{\bl}B] -\prob[B\geq 0] +
\frac{\pdff_B'(0)}{2\bl} \right|\leq k_1.\IEEEeqnarraynumspace
\end{IEEEeqnarray}
%
\end{lemma}
\ifthenelse{\boolean{confversion}}
{}
{\begin{IEEEproof}
See Appendix~\ref{app:proof-lem-expectation-phi}.
\end{IEEEproof}
}

From~\eqref{eq:int2} and~\eqref{eq:int3}, and
recalling~\eqref{eq:epsilon-capacity} we may expect that for
a quasi-static fading channel $R^*(n, \epsilon)$ satisfies
\begin{IEEEeqnarray}{rCl}
 R^*(n, \epsilon) = C_\epsilon + 0\cdot {1\over \sqrt{n}} + \text{smaller-order
terms}\,.
\end{IEEEeqnarray}

This intuitive reasoning turns out to be correct as the following result demonstrates.

\begin{thm}
\label{thm:dispersion-simo}
Assume that the channel gain $G=\|\randvech\|^2$ has a
twice continuously differentiable pdf and that $C_\epsilon$ is a point of growth of the
capacity-outage function~\eqref{eq:def-cadist}, i.e., $F'_C(C_\error)>0$.
%
%
Then, the maximal achievable rates satisfy
\ifthenelse{\boolean{confversion}}
{
\begin{IEEEeqnarray}{rCl}
\label{eq:thm-R-star-expansion}
\big\{\Rnocsit^\ast(\bl,\error),\, \Rcsirt^\ast(\bl,\error)\big\}&=&  C_\error + \bigO\lefto(\log(\bl) /\bl\right).
\end{IEEEeqnarray}
}{\begin{IEEEeqnarray}{rCl}
\label{eq:thm-R-star-expansion-1}
\Rnocsit^\ast(\bl,\error) &=&  C_\error + \bigO\lefto(\frac{\log\bl }{\bl}\right)\\
\label{eq:thm-R-star-expansion-2}
\Rcsirt^\ast(\bl,\error) &=&  C_\error + \bigO\lefto(\frac{\log \bl}{\bl}\right).
\end{IEEEeqnarray}}
Hence, the $\error$-dispersion is zero for both the no-CSI and the CSIRT case:
\begin{equation}
\Vnocsit = \Vcsit = 0\,,\qquad \error\in(0,1)\backslash\{1/2\}\,.
\end{equation}
\end{thm}
\ifthenelse{\boolean{confversion}}
{
\begin{IEEEproof}
The proof is outlined in Appendix~\ref{sec:proof-theorem}.
\end{IEEEproof} }
{
\begin{IEEEproof}
See Appendix~\ref{sec:proof-theorem}.
\end{IEEEproof}
}

\setlength{\baselineskip}{1.16em}
{The assumptions on the channel gain are satisfied by the probability distributions commonly used to model fading, such as Rayleigh, Rician, and Nakagami.
However, the standard AWGN channel, which can be seen as a quasi-static fading channel with fading distribution equal to a step function centered at one, does not meet these assumptions and in fact has positive dispersion~\cite[Thm.~54]{polyanskiy10-05}.
%

Note that, as the fading distribution approaches a step function, the higher-order terms in the expansion~\ifthenelse{\boolean{confversion}}{\eqref{eq:thm-R-star-expansion}}{\eqref{eq:thm-R-star-expansion-1} and~\eqref{eq:thm-R-star-expansion-2}} become more dominant, and zero dispersion does not necessarily imply fast convergence to capacity. 
Consider for example a single-input single-output Rician fading with Rician factor $K$.
\ifthenelse{\boolean{confversion}}
{For $\error<1/2$, one can refine~\ifthenelse{\boolean{confversion}}{\eqref{eq:thm-R-star-expansion}}{\eqref{eq:thm-R-star-expansion-1} and~\eqref{eq:thm-R-star-expansion-2}} and show that
\begin{IEEEeqnarray}{rCl}
&&C_\error-\frac{\log\bl}{\bl} +\frac{c_1 \sqrt{K} + c_2}{\bl} + \littleo\lefto(\frac{1}{\bl}\right) \leq \Rnocsit^\ast(\bl,\error)\notag\\
&&\quad\quad\leq \Rcsirt^\ast(\bl,\error) \leq C_\error + \frac{\log\bl}{\bl} + \frac{\tilde{c}_1 \sqrt{K}+\tilde{c}_2}{\bl} + \littleo\lefto(\frac{1}{\bl}\right)\IEEEeqnarraynumspace
\label{eq:eg-rician-bounds}
\end{IEEEeqnarray}
where $c_1$, $c_2$, $\tilde{c}_1$ and $\tilde{c}_2$ are finite constants with $c_1<0$ and $\tilde{c}_1<0$. As $K$ increases and the fading distribution converges to a step function, the third term in both the upper and lower bounds in~\eqref{eq:eg-rician-bounds} becomes increasingly large in absolute value.}
{The pdf of $G$ is
\begin{IEEEeqnarray}{rCl}
\label{eq:pdf-rician-g}
\pdff_G(g) = (K+1)e^{-K-(K+1)g}I_0(2\sqrt{K(K+1)g})
\end{IEEEeqnarray}
where $I_0(\cdot)$ denotes the zero-th order modified Bessel function of the first kind. It follows from~\eqref{eq:conv-Rrt-gamma-n} and~\eqref{eq:coh-final-step} in Appendix~\ref{sec:proof-theorem} that
\begin{IEEEeqnarray}{rCl}
\Rcsirt^\ast(\bl,\error) &\leq& C_\error + \frac{\log\bl}{\bl} + \frac{\funcU(C_\error)+2}{2 \cadist'(C_\error)}\cdot\frac{1}{\bl} + \littleo\lefto(\frac{1}{\bl}\right)\\
&=& C_\error + \frac{\log\bl}{\bl} + \frac{\snr e^{-C_\error}}{\bl \pdff_G(g_0)} - \frac{1+e^{-2C_\error}}{2\bl} \notag\\
&&-\,\frac{\snr(e^{C_\error} -e^{-C_\error})}{2\bl}\frac{\pdff'_G(g_0)}{\pdff_G(g_0)} + \littleo\lefto(\frac{1}{\bl}\right)\label{eq:conv-rician}
\end{IEEEeqnarray}
where $\funcU(\cdot)$ is defined in~\eqref{eq:def-funcU} and $g_0\define(e^{C_\error}-1)/\snr$. Without loss of generality, we assume that $\error<1/2$. Then, it can be shown that there exists a constant $K_\error>0$ such that for all $K\geq K_\error$,
\begin{IEEEeqnarray}{rCl}
\label{eq:bound-g0-rician}
1-\frac{c_1}{\sqrt{K}} \leq g_0\leq 1-\frac{c_2}{\sqrt{K}}
\end{IEEEeqnarray}
for some $c_1\geq c_2>0$. The ratio ${\pdff'_G(g_0)}/{\pdff_G(g_0)}$ can be computed as
\begin{IEEEeqnarray}{rCl}
\label{eq:ratio-pdf-G}
\frac{\pdff'_G(g_0)}{\pdff_G(g_0)} = \frac{\sqrt{K(K+1)}}{\sqrt{g_0}}\frac{I_1(2\sqrt{K(K+1)g_0})}{I_0(2\sqrt{K(K+1)g_0})}-K-1.\IEEEeqnarraynumspace
\end{IEEEeqnarray}
By using in~\eqref{eq:ratio-pdf-G} the bound (see, e.g., \cite[Eq.~(1.12)]{laforgia10})
\begin{IEEEeqnarray}{rCl}
 \frac{I_1(x)}{I_0(x)} \geq \frac{\sqrt{x^2+1}-1}{x}
\end{IEEEeqnarray}
and the inequality~\eqref{eq:bound-g0-rician}, we get
\begin{IEEEeqnarray}{rCl}
\label{eq:bound-ratio-pdf-G}
\frac{\pdff'_G(g_0)}{\pdff_G(g_0)} \geq c_3\sqrt{K} +c_4
\end{IEEEeqnarray}
for some constants $c_3>0$ and $c_4$.
As $\pdff_G(g)$ is unimodal, it can be bounded from below as follows:
\begin{IEEEeqnarray}{rCl}
\label{eq:bound-pdf-G}
f_G(g_0) \geq \frac{\error}{g_0} \geq  \error.
\end{IEEEeqnarray}
Substituting \eqref{eq:bound-ratio-pdf-G} and~\eqref{eq:bound-pdf-G} into~\eqref{eq:conv-rician}, we get
\begin{IEEEeqnarray}{rCl}
\label{eq:ub-R-rician}
\Rcsirt^\ast(\bl,\error) &\leq& C_\error + \frac{\log\bl}{\bl} + \frac{c_5\sqrt{K}+c_6}{\bl} + \littleo\lefto(\frac{1}{\bl}\right)
\end{IEEEeqnarray}
where $c_5<0$ and $c_6$ are finite constants.
%
 Following similar steps, one can also establish that
\begin{IEEEeqnarray}{rCl}
\label{eq:lb-R-rician}
\Rnocsit^\ast(\bl,\error) &\geq& C_\error - \frac{\log\bl}{\bl} + \frac{\tilde{c}_1\sqrt{K} + \tilde{c}_2}{\bl} + \littleo\lefto(\frac{1}{\bl}\right)
\end{IEEEeqnarray}
for some finite $\tilde{c}_1<0$ and $\tilde{c}_2$. We see from~\eqref{eq:ub-R-rician} and~\eqref{eq:lb-R-rician} that, as $K$ increases and the fading distribution converges to a step function, the third term in the RHS of~\eqref{eq:ub-R-rician} and~\eqref{eq:lb-R-rician} becomes increasingly large in absolute value.
}
}

\subsection{Numerical Results}
\label{sec:numerical}
\ifthenelse{\boolean{confversion}}
{\begin{figure}[t]
	\centering
		\includegraphics[scale=0.75]{rician.pdf}
\caption{Bounds for the quasi-static SIMO Rician-fading channel with $K$-factor equal to $20$ dB, two receive antennas, $\text{SNR}=-1.55 $ dB, and \mbox{$\epsilon=10^{-3}$}.\label{fig:gamma_distribution}}
\vspace{-6mm}
\end{figure}}
{

}

Fig.~\ref{fig:gamma_distribution} shows the achievability bound~\eqref{eq:lb-numcode-simo_final} and the converse bound (\ref{eq:thm-converse-rcsit}) for a quasi-static SIMO fading channel with two receive antennas.
The channel between the transmit antenna and each of the two receive antennas is Rician-distributed with $K$-factor equal to $20$ dB.
The two channels are assumed to be independent.
We set $\error=10^{-3}$ and choose $\snr=-1.55$~dB so that $C_\error =1$ bit$/$channel use.
For reference, we also plotted a lower bound on $\Rcsirt^\ast(\bl,\error)$ obtained by using the~$\kappa\beta$ bound~\cite[Thm.~25]{polyanskiy10-05} and assuming CSIR.\footnote{Specifically, we took $\inset=\inset_{\bl}$ with $\inset_{\bl}$  defined in (\ref{eq:equal_power_constraint}), and $\outdist_{\randmatY\randvech}=\indist_{\randvech}\outdist_{\randmatY\given \randvech}$ with $\outdist_{\randmatY\given \randvech}$ defined in (\ref{eq:converse-Qchannel-def}).}
%
Fig.~\ref{fig:gamma_distribution} shows also the approximation~\eqref{eq:approx_R_introduction} for $R^\ast(\bl,\error)$ corresponding to an AWGN channel with $C =1$ bit$/$channel use. Note that
we replaced the term $\bigO(\log(\bl)/\bl)$ in (\ref{eq:approx_R_introduction}) with $\log(\bl)/(2\bl)$ (see \cite[Eq.~(296)]{polyanskiy10-05}).\footnote{The validity of the approximation~\cite[Eq.~(296)]{polyanskiy10-05} is numerically verified in~\cite{polyanskiy10-05} for a real AWGN channel.
Since a complex AWGN channel can be treated as two real AWGN channels with the same SNR,  the approximation~\cite[Eq.~(296)]{polyanskiy10-05} with $C=\log(1+\snr)$ and $V=\frac{\snr^2+2\snr}{(1+\snr)^2}$ is accurate for the complex case~\cite[Thm.~78]{polyanskiy10}.}
The blocklength required to achieve $90\%$ of the $\error$-capacity of the quasi-static fading channel is in the range $[120,320]$ for the CSIRT case and in the range $[120,480]$ for the no-CSI case.
For the AWGN channel, this number is approximately~$1420$. Hence, for the parameters chosen in Fig.~\ref{fig:gamma_distribution}, the prediction (based on zero dispersion) of fast convergence to capacity is validated.

%
%

\section*{Acknowledgements}
Initial versions of these results were discussed by Y.~Polyanskiy with Profs. H.~V.~Poor and S. Verd\'u, whose support and comments are kindly acknowledged.

\appendices

\ifthenelse{\boolean{confversion}}{\section{} }
{\section{Proof of Theorem~\ref{thm:converse}}}
\label{sec:app-proof-converse}
For the channel~\eqref{eq:channel_io_new} with CSIRT, the input is the pair $(\inseqrand,\randvech)$, and the output is the pair $(\outseqrand, \randvech)$.
Note that the encoder induces a distribution $\indist_{\inseqrand \given \randvech}$ on $\inseqrand$ and is necessarily randomized, since $\randvech$ is independent of the message~$\msg$.
Denote by $\equalR^\ast(\bl ,\error)$ the maximal achievable rate under the constraint that each codeword $\vecc_j{(\vech)}$ satisfies the power constraint (\ref{eq:power-constraint-csirt}) with equality, namely,
\ifthenelse{\boolean{confversion}}
{
$\vecc_j{(\vech)}\in\inset_{\bl}$
}
{$\vecc_j{(\vech)}$ belongs to the set $\inset_\bl$ defined in~\eqref{eq:equal_power_constraint}}
for $j=1,\ldots,\NumCode$ and for all $\vech\in\complexset^r$.
%
%
Then by \cite[Lem.~39]{polyanskiy10-05}, 
\begin{IEEEeqnarray}{rCl}
\label{eq:equal-power-max-power-relation}
\Rcsirt^\ast(\bl-1,\error) \leq  \frac{n}{n-1}\equalR^\ast(\bl,\error).
\end{IEEEeqnarray}
\ifthenelse{\boolean{confversion}}
{We next establish an upper bound on $\equalR^\ast(\bl,\error)$ using the meta-converse theorem~\cite[Thm.~26]{polyanskiy10-05}.
}
{We next establish an upper bound on $\equalR^\ast(\bl,\error)$.
Henceforth, $\inseq$ is assumed to belong to $\inset_\bl$.
To upper-bound $\equalR^\ast(\bl,\error)$, we use the meta-converse theorem~\cite[Thm.~26]{polyanskiy10-05}.}
As \emph{auxiliary} channel $\outdist_{\outseqrand \randvech\given \inseqrand\randvech}$, we take a channel that passes $\randvech$ unchanged and generates $\outseqrand$ according to the following distribution
\begin{IEEEeqnarray}{rCl}
\label{eq:converse-Qchannel-def}
\outdist_{ \outseqrand \given \randvech=\vech,\inseqrand=\vecx} = \prod\limits_{j=1}^{\bl} \jpg(\veczero, \matI_{\rxant}+\snr\vech\herm{\vech}).
\end{IEEEeqnarray}
In particular, $\outseqrand$ and $\inseqrand$ are conditionally independent given~$\randvech$.
Since $\randvech$ and the message $\msg$ are independent,  $\outseqrand$ and $\msg$ are independent under the auxiliary $\outdist$-channel.
Hence, the average error probability $\error'$ under the auxiliary $\outdist$-channel is bounded as
\ifthenelse{\boolean{confversion}}
{$\error'\geq 1- 1/\NumCode$.
}
{
\begin{IEEEeqnarray}{rCl}
\error'\geq 1- \frac{1}{\NumCode}.
\end{IEEEeqnarray}}
%
%
Then,~\cite[Thm.~26]{polyanskiy10-05}
\begin{IEEEeqnarray}{rCl}
\bl \equalR^\ast(\bl,\error)
\leq\! \sup_{\indist_{\inseqrand\given\randvech}}\!\log\lefto(\frac{1}{\beta_{1-\error}(\indist_{\inseqrand\outseqrand\randvech}, \indist_{\randvech} \indist_{\inseqrand\given \randvech} \outdist_{\outseqrand\given \randvech} )}\right)\,\,\,\,\quad\label{eq:general-converse-R}
\end{IEEEeqnarray}
\ifthenelse{\boolean{confversion}}
{where $\beta_{1-\error}(\cdot,\cdot)$ is defined in~\cite[Eq.~(100)]{polyanskiy10-05},}
{where $\beta_{1-\error}(\cdot,\cdot)$ is defined in~\eqref{eq:def-beta},} and the supremum is over all conditional distributions $\indist_{\inseqrand \given \randvech}$ supported on $\inset_\bl$.
%
%
We next note that, by the spherical symmetry of $\inset_\bl$ and of (\ref{eq:converse-Qchannel-def}), the function $\beta_\alpha (\indist_{\outseqrand \given \inseqrand= \inseq,\randvech=\vech} , \outdist_{\outseqrand\given \randvech=\vech})$ does not depend on $\inseq \in \inset_{\bl}$. By \cite[Lem.~29]{polyanskiy10-05}, this implies
\begin{IEEEeqnarray}{rCl}
\IEEEeqnarraymulticol{3}{l}{\beta_\alpha ( \indist_{\inseqrand \outseqrand \given \randvech=\vech} , \indist_{\inseqrand\given \randvech =\vech} \outdist_{\outseqrand\given \randvech=\vech})}\notag\\
\,\, &=&\beta_\alpha (\indist_{\outseqrand \given \inseqrand= \inseq_0,\randvech=\vech} , \outdist_{\outseqrand\given \randvech=\vech})\label{eq:conditional-beta}
\end{IEEEeqnarray}
(with $\inseq_0$ defined in~\eqref{eq:vec0}) for every $\indist_{\inseqrand\given \randvech=\vech}$ supported on~$\inset_\bl$, every $\vech\in\complexset^r$, and every $\alpha$.
\ifthenelse{\boolean{confversion}}
{Following similar steps as in the proof of \cite[Lem.~29]{polyanskiy10-05} and using (\ref{eq:conditional-beta}), we conclude that
\begin{IEEEeqnarray}{rCl}
\IEEEeqnarraymulticol{3}{l}{
\beta_{1-\error}(\indist_{\inseqrand\outseqrand\randvech},\indist_{\randvech}  \indist_{\inseqrand\given \randvech} \outdist_{\outseqrand\given \randvech} )}\notag\\
 & = &\beta_{1-\error} (\indist_{\randvech} \indist_{\outseqrand \given \inseqrand= \inseq_0,\randvech} , \indist_{\randvech}  \outdist_{\outseqrand\given \randvech})\label{eq:lb-beta-joint}
\end{IEEEeqnarray}
for every $\indist_{\inseqrand\given \randvech}$ supported on $\inset_\bl$.
}
{Consider the optimal test $Z$ for $\indist_{\inseqrand\outseqrand\randvech}$ versus $\indist_{\randvech}  \indist_{\inseqrand\given \randvech} \outdist_{\outseqrand\given \randvech} $ under the constraint that
\begin{equation}
\label{eq:def-alpha-h}
\indist_{\inseqrand\outseqrand\randvech}[Z=1]=\int \underbrace{\indist_{\inseqrand\outseqrand\given\randvech=\vech}[Z=1]}_{\define \alpha(\vech)}d\indist_{\randvech}(\vech)\geq 1-\error.
\end{equation}
We have that
\begin{IEEEeqnarray}{rCl}
\IEEEeqnarraymulticol{3}{l}{\beta_{1-\error} (\indist_{\inseqrand\outseqrand\randvech},\indist_{\randvech}  \indist_{\inseqrand\given \randvech} \outdist_{\outseqrand\given \randvech} )}\notag\\
\,&=& \int \indist_{\inseqrand\given \randvech= \vech}\outdist_{\outseqrand\given \randvech=\vech}[Z=1] d\indist_{\randvech}(\vech)\\
&\geq & \int \beta_{\alpha(\vech)} ( \indist_{\inseqrand \outseqrand \given \randvech=\vech} , \indist_{\inseqrand\given \randvech =\vech} \outdist_{\outseqrand\given \randvech=\vech}) d\indist_{\randvech}(\vech) \,\,\,\,\\
&=&\int \beta_{\alpha(\vech)} ( \indist_{\outseqrand \given \inseqrand= \inseq_0,\randvech=\vech} , \outdist_{\outseqrand\given \randvech=\vech}) d\indist_{\randvech}(\vech) \label{eq:lower-bound-beta-first-step}
\end{IEEEeqnarray}
where (\ref{eq:lower-bound-beta-first-step}) follows from (\ref{eq:conditional-beta}). Fix an arbitrary $\vech\in\complexset^{\rxant}$, and let $Z^{*}_{\vech}$ be an optimal test between $\indist_{\outseqrand \given \inseqrand = \inseq_0, \randvech=\vech}$ and $ \outdist_{\outseqrand\given \randvech=\vech}$, i.e., a test satisfying
\begin{IEEEeqnarray}{rCl}
\indist_{\outseqrand \given \inseqrand = \inseq_0,\randvech=\vech}[Z^{*}_{\vech}=1]&\geq& \alpha(\vech)
\end{IEEEeqnarray}
and
\begin{IEEEeqnarray}{rCl}
\outdist_{\outseqrand\given \randvech=\vech}[Z^{*}_{\vech}=1] &=& \beta_{\alpha(\vech)} ( \indist_{\outseqrand \given \inseqrand= \inseq_0,\randvech=\vech} , \outdist_{\outseqrand\given \randvech=\vech}).\IEEEeqnarraynumspace \label{eq:def-q-beta}
\end{IEEEeqnarray}
Then $Z^{*}_{\randvech}$ is a test between $\indist_{\randvech}\indist_{\outseqrand \given \inseqrand = \inseq_0, \randvech}$ and $ \indist_{\randvech}\outdist_{\outseqrand\given \randvech}$.
Moreover,
\begin{IEEEeqnarray}{rCl}
\int \indist_{\outseqrand \given \inseqrand = \inseq_0,\randvech=\vech}[Z^{*}_{\vech}=1] d\indist_{\randvech}(\vech) &\geq& \int  \alpha(\vech) d\indist_{\randvech}(\vech) \IEEEeqnarraynumspace\\
&\geq& 1-\error \label{eq:average-alpha-lb}
\end{IEEEeqnarray}
where (\ref{eq:average-alpha-lb}) follows from (\ref{eq:def-alpha-h}). Consequently,
\begin{IEEEeqnarray}{rCl}
\IEEEeqnarraymulticol{3}{l}{
\int\beta_{\alpha(\vech)} ( \indist_{\outseqrand \given \inseqrand= \inseq_0,\randvech=\vech} , \outdist_{\outseqrand\given \randvech=\vech})d\indist_{\randvech}(\vech)}\notag\\
\,\,&=&\int \outdist_{\outseqrand\given \randvech=\vech}[Z^{*}_{\vech}=1]d\indist_{\randvech}(\vech) \label{eq:lower-bound-beta-second-step-mid}\\
&\geq &\beta_{1-\error}( \indist_{\randvech} \indist_{\outseqrand \given \inseqrand= \inseq_0,\randvech} , \indist_{\randvech}  \outdist_{\outseqrand\given \randvech})\label{eq:lower-bound-beta-second-step}
\end{IEEEeqnarray}
where (\ref{eq:lower-bound-beta-second-step-mid}) follows from (\ref{eq:def-q-beta}), and (\ref{eq:lower-bound-beta-second-step}) follows by the definition of $\beta_{1-\error}(\cdot,\cdot)$ and by (\ref{eq:average-alpha-lb}). Substituting (\ref{eq:lower-bound-beta-second-step}) into (\ref{eq:lower-bound-beta-first-step}), we obtain that
\begin{IEEEeqnarray}{rCl}
\IEEEeqnarraymulticol{3}{l}{
\beta_{1-\error}(\indist_{\inseqrand\outseqrand\randvech},\indist_{\randvech}  \indist_{\inseqrand\given \randvech} \outdist_{\outseqrand\given \randvech} )}\notag\\
 &\geq&\beta_{1-\error} (\indist_{\randvech} \indist_{\outseqrand \given \inseqrand= \inseq_0,\randvech} , \indist_{\randvech}  \outdist_{\outseqrand\given \randvech})\label{eq:lb-beta-joint}
\end{IEEEeqnarray}
for every $\indist_{\inseqrand\given \randvech}$ supported on $\inset_\bl$.
  It can be shown that (\ref{eq:lb-beta-joint}) holds, in fact, with equality.
}

In the following,
to shorten notation, we define 
\begin{IEEEeqnarray}{rCl}
\indist_0 \define \indist_{\randvech} \indist_{\outseqrand \given \inseqrand= \pickcwd,\randvech},\quad
 \outdist_0 \define \indist_{\randvech}  \outdist_{\outseqrand\given \randvech}.
\end{IEEEeqnarray}
%
Using this notation, (\ref{eq:general-converse-R}) becomes
\begin{IEEEeqnarray}{rCl}
\label{eq:converse-R}
\bl \equalR^\ast(\bl,\error) \leq - \log \beta_{1-\error}( \indist_0, \outdist_0).
\end{IEEEeqnarray}
%
%
%
Let
\ifthenelse{\boolean{confversion}}{$r(\pickcwd;\outseqrand\randvech)  \define \log \lefto(d\indist_{0}/d\outdist_{0}\right)$.}
{\begin{IEEEeqnarray}{rCl}%
r(\pickcwd;\outseqrand\randvech)  \define \log {d\indist_{0}\over d\outdist_{0}}.
\end{IEEEeqnarray}}
By the Neyman-Pearson lemma (see for example~\cite[p.~23]{poor94a}),
\begin{align}\label{eq:neyman_pearson_one}
  \beta_{1-\error}( \indist_0, \outdist_0) = \outdist_{0}\bigl[r(\pickcwd;\outseqrand\randvech) \geq \bl \gamma_{\bl} \bigr]
\end{align}
where $\gamma_{\bl}$ is the solution of
\ifthenelse{\boolean{confversion}}{$\indist_{0}\bigl[r(\pickcwd;\outseqrand\randvech) \leq \bl \gamma_{\bl}\bigr]=\error$.}
{\begin{align}
  \indist_{0}\bigl[r(\pickcwd;\outseqrand\randvech) \leq \bl \gamma_{\bl}\bigr]=\error.
\end{align}
}
We conclude the proof by noting that, under $\outdist_{0}$, the random variable $r(\pickcwd;\outseqrand\randvech)$ has the same distribution as $L_\bl $ in~\eqref{eq:info_density_simo_alt}, and under $\indist_{0}$, it has the same distribution as $S_\bl$ in~\eqref{eq:info_density_simo}.

\ifthenelse{\boolean{confversion}}{\section{} }
{
\vspace{0.3cm}
\section{Proof of Corollary~\ref{cor:actual_achievability_bound}}}
\label{app:corollary}

Due to spherical symmetry and to the assumption that \mbox{$\vecx\in \inset_\bl$}, the term $\indist_{\randmatY\given \randvecx = \vecx}[\cos^2\theta(\vecx, \randmatY)\geq 1-\gamma_\bl]$ on the LHS of~\eqref{eq:threshold_tester_Z}, does not depend on $\vecx$.
Hence, we can set $\vecx=\vecx_0$.

We next evaluate $\sup_{\vecx\in \inset_\bl}\outdist_{\randmatY}[\angletest_{\vecx}(\randmatY)=1]$ for the Gaussian distribution $\outdist_{\randmatY}$ in~\eqref{eq:Gaussian_output_distribution}.
Under $\outdist_{\randmatY}$, the random subspace spanned by the columns of $\randmatY$ is $\rxant$-dimensional with probability one, and is uniformly distributed on the Grassmann manifold of $\rxant$-planes in $\complexset^{\bl}$~\cite[Sec.~6]{james54}.
If we take $\randveca\sim\outdist_{\randveca} = \jpg(\mathbf{0},\matI_\bl)$ to be independent of $\randmatY\sim\outdist_{\randmatY}$, then for every $\vecx \in \inset_\bl$ and every $\matY\in \complexset^{\bl\times\rxant}$ with full column rank
\begin{IEEEeqnarray}{rCl}
\outdist_{\randmatY}[\angletest_{\vecx}(\randmatY)=1] &=& \outdist_{\randmatY, \randveca}[\angletest_{\randveca}(\randmatY)=1]\label{eq:equal-probability-a}\\
&= &\outdist_{\randveca}[\angletest_{\randveca}(\matY)=1]\label{eq:equal-probability-c}.
\end{IEEEeqnarray}
In~(\ref{eq:equal-probability-a}) we used that $\outdist_{\randmatY}[\angletest_{\vecx}(\randmatY)=1]$ does not depend on $\vecx$;
(\ref{eq:equal-probability-c}) holds because $\outdist_{\randveca}$ is isotropic.

To compute the RHS of~\eqref{eq:equal-probability-c}, we will choose for simplicity
\begin{IEEEeqnarray}{rCl}
\matY =\left[
             \begin{array}{c}
               \matI_{\rxant} \\
               \mathbf{0}_{(\bl-\rxant)\times \rxant} \\
             \end{array}
           \right].
\end{IEEEeqnarray}
The columns of $\matY$ are orthonormal.
Hence, by (\ref{eq:ortho-cos-theta}) and (\ref{eq:threshold_tester_Z})
\ifthenelse{\boolean{confversion}}
{
\begin{IEEEeqnarray}{rCl}
\outdist_{\randveca}[\angletest_{\randveca}(\matY)=1] &=&\outdist_{\randveca} \lefto[ \|\herm{\randveca}\matY \|^2 /\|\randveca\|^2 \geq 1 - \gamma_\bl\right]\\
 &=& \outdist_{\randveca} \lefto[\frac{\sum\nolimits_{i=r+1}^{\bl}|A_i|^2}{\sum\nolimits_{i=1}^{\bl}|A_i|^2}\leq\gamma_\bl\right]\label{eq:comput_Q-gamma-ratio-beta}
\end{IEEEeqnarray}
}
{
\begin{IEEEeqnarray}{rCl}
\outdist_{\randveca}[\angletest_{\randveca}(\matY)=1] &=&\outdist_{\randveca} \lefto[ \frac{ \|\herm{\randveca}\matY \|^2 }{\|\randveca\|^2} \geq 1 - \gamma_\bl\right]\\
 &=& \outdist_{\randveca} \lefto[\frac{\sum\nolimits_{i=r+1}^{\bl}|A_i|^2}{\sum\nolimits_{i=1}^{\bl}|A_i|^2}\leq\gamma_\bl\right]\label{eq:comput_Q-gamma-ratio-beta}
\end{IEEEeqnarray}
}
where $A_i\sim \jpg(0,1)$ is the $i$th entry of $\randveca$. Observe that the ratio
\ifthenelse{\boolean{confversion}}
{ $\left(\sum\nolimits_{i=r+1}^{\bl}|A_i|^2\right)/\left(\sum\nolimits_{i=1}^{\bl}|A_i|^2\right)$
}{\begin{equation}
  {\sum\nolimits_{i=r+1}^{\bl}|A_i|^2 \over \sum\nolimits_{i=1}^{\bl}|A_i|^2}
\end{equation}}
is $\mathrm{Beta}(\bl-\rxant,\rxant)$-distributed~\cite[Ch.~25.2]{johnson95-2}.

To conclude the proof, we need to compute $\tilde{\kappa}_{\tau}(\inset_\bl, \outdist_{\randmatY})$.
If we replace the constraint \ref{item:constraint1}) in (\ref{eq:def_kappa_tilde}) by the less stringent constraint that
\begin{IEEEeqnarray}{rCl}
\label{eq:refined-kappa-beta-constraint1}
\indist_{\randmatY}[Z(\randmatY)=1] = \Ex{\indist^{(\text{unif})}_{\randvecx}}{\indist_{\randmatY\given \randvecx}[Z(\randmatY)=1]} \geq \tau
\end{IEEEeqnarray}
 with $\indist_{\randmatY}$ being the output distribution induced by the uniform input distribution $\indist^{(\text{unif})}_{\randvecx}$ on $\inset_\bl$, we get an infimum in (\ref{eq:def_kappa_tilde}), which we denote by $\bar{\kappa}_\tau$, that is no larger than $\tilde{\kappa}_\tau(\inset_\bl, \outdist_{\randmatY})$.
  Because both $\outdist_{\randmatY}$ and the output distribution $\indist_{\randmatY}$ induced by $\indist^{(\text{unif})}_{\randvecx}$ are isotropic, we conclude that
\begin{IEEEeqnarray}{rCl}
  \indist_{\randmatY}[Z(\randmatY)=1]=\outdist_{\randmatY}[ Z(\randmatY)=1 ]\geq \tau
\end{IEEEeqnarray}
for all tests $Z(\randmatY)$ that satisfy (\ref{eq:refined-kappa-beta-constraint1}) and the constraint \ref{item:constraint2}) in (\ref{eq:def_kappa_tilde}). Therefore,
\ifthenelse{\boolean{confversion}}
{$\tilde{\kappa}_{\tau}(\inset_\bl, \outdist_{\randmatY}) \geq \bar{\kappa}_\tau = \tau$.
}{
\begin{IEEEeqnarray}{rCl}
\label{eq:kappa-tau-tilde}
\tilde{\kappa}_{\tau}(\inset_\bl, \outdist_{\randmatY}) \geq \bar{\kappa}_\tau = \tau.
\end{IEEEeqnarray}
}

\ifthenelse{\boolean{confversion}}
{
}
{

 \section{Proof of Lemma~\ref{lem:expectation-phi}}
\label{app:proof-lem-expectation-phi}
By assumption, there exist $\delta>0$ and $k_2<\infty$, such that
 \begin{IEEEeqnarray}{rCl}
 \max\{|\pdff_{B}(t)|,|\pdff_{B}'(t)|,|\pdff_{B}''(t)| \}\leq k_2
 \end{IEEEeqnarray}
 for all $t \in (-\delta,\delta)$.
Let $\cdfF_B$ be the cdf of $B$.
We write
\begin{IEEEeqnarray}{rCl}
\prob[A\leq \sqrt{\bl}B] &=& \int\nolimits_{|a|\geq\delta\sqrt{\bl}}\prob[B \geq a/\sqrt{\bl}] d\indist_A \notag\\
&&+ \int\nolimits_{|a|< \delta\sqrt{\bl}}\prob[B \geq a/\sqrt{\bl}]d\indist_A\\
&=& \int\nolimits_{|a|\geq \delta\sqrt{\bl}}\prob[B \geq a/\sqrt{\bl}] d\indist_A \notag\\
&&+\int\nolimits_{|a|< \delta\sqrt{\bl}} (1-\cdfF_B(a/\sqrt{\bl}))d\indist_A  \label{eq:app-prob-x-leq-u}.
\end{IEEEeqnarray}
We next evaluate the two terms on the RHS of (\ref{eq:app-prob-x-leq-u}). For the first term, we have that
\begin{IEEEeqnarray}{rCl}
 \int\nolimits_{|a|\geq\delta\sqrt{\bl}}\!\!\prob[B \geq a/\sqrt{\bl}] d\indist_A &\leq& \int\nolimits_{|a|\geq \delta\sqrt{\bl}} d\indist_A \\
   &{\leq}& \frac{\Ex{}{|A|^3}}{\delta^3\bl^{3/2}} \label{eq:lemma_outer_term}
 \end{IEEEeqnarray}
 where (\ref{eq:lemma_outer_term}) follows from Markov's inequality.
%
To compute the second term on the RHS of (\ref{eq:app-prob-x-leq-u}), we note that, by Taylor's theorem~\cite[Thm.~5.15]{rudin76}, for all $a\in(-\delta \sqrt{\bl},\delta\sqrt{\bl})$,
\begin{IEEEeqnarray}{rCl}
\IEEEeqnarraymulticol{3}{l}{
\cdfF_B(a/\sqrt{\bl})}\notag\\ \,\,\,\,&=& \cdfF_B(0) + \pdff_B(0)\frac{a}{\sqrt{\bl}} + \frac{\pdff'_B(0)}{2}\frac{a^2}{\bl}
                + \frac{\pdff''_B(a_0)}{6}\frac{a^3}{\bl^{3/2}}\IEEEeqnarraynumspace
\end{IEEEeqnarray}
for some $a_0 \in (0, a/\sqrt{\bl})$.
Averaging over $A$, we get
\begin{IEEEeqnarray}{rCl}
\IEEEeqnarraymulticol{3}{l}{
\int\nolimits_{|a|<\delta\sqrt{\bl}}\cdfF_B(a/\sqrt{\bl})  d\indist_A } \notag\\
\;\; &=& \cdfF_B(0)(1-\underbrace{\prob[|a|\geq \delta\sqrt{\bl}]}_{\leq \delta^{-3}\Ex{}{|A|^3} \bl^{-3/2}})\notag\\
&& +\; \frac{\pdff_B(0)}{\sqrt{\bl}}\underbrace{\Ex{}{A \cdot
\indfun{|A| < \delta \sqrt{\bl}}}}_{\define c_1(\bl)} \notag\\
&& +\; \frac{\pdff'_B(0)}{2\bl}\Big(\Ex{}{A^2}-\underbrace{\Ex{}{A^2 \cdot \indfun{|A|\geq \delta\sqrt{\bl}}}}_{\define c_2(\bl)}\Big) \notag\\
&& +\; \underbrace{\Ex{}{\frac{A^3 \pdff_{B}''(A_0)}{6\bl^{3/2}} \cdot \indfun{|A|<\delta\sqrt{\bl}}}}_{\define c_3(n)}.
\label{eq:taylor_expectation}\IEEEeqnarraynumspace
\end{IEEEeqnarray}
The term $c_1(\bl)$ can be bounded as
\begin{IEEEeqnarray}{rCl}
\left|c_1(\bl)\right| &=& \big|\underbrace{\Ex{}{A}}_{=0}{} - \Ex{}{A\cdot \indfun{|A|\geq \delta \sqrt{\bl}}}\big|\\
&=& \big|\Ex{}{A \cdot \indfun{|A|\geq \delta \sqrt{\bl}}}\big|\\
 &\leq& \Ex{}{|A| \cdot \indfun{|A|\geq \delta \sqrt{\bl}}}\\
 &=&\frac{1}{\delta^2\bl} \Ex{}{\delta^2 \bl |A| \cdot \indfun{ |A|\geq \delta \sqrt{\bl} } }\\
 &\leq & \frac{1}{\delta^2\bl} \Ex{}{ |A|^3 \cdot \indfun{|A|\geq \delta \sqrt{\bl}}}\label{eq:app-bound-x-u-c1}.
\end{IEEEeqnarray}
The term $c_2(\bl)$ can be bounded as
\begin{IEEEeqnarray}{rCl}
|c_2(\bl)|&=& \frac{1}{\delta\sqrt{\bl}} \Ex{}{\delta\sqrt{\bl}\cdot |A|^2 \cdot \indfun{|A|\geq \delta \sqrt{\bl}} }\\
 &\leq & \frac{1}{\delta\sqrt{\bl}} \Ex{}{ |A|^3 \cdot \indfun{|A|\geq \delta \sqrt{\bl}} }.\label{eq:app-bound-x-u-c2}
\end{IEEEeqnarray}
Finally, $c_3(\bl)$ can be bounded as
\begin{IEEEeqnarray}{rCl}
|c_3(\bl)|&\leq& \Ex{}{\frac{|A|^3 |\pdff_{B}''(A_0)| }{6\bl^{3/2}} \cdot \indfun{|A|<\delta\sqrt{\bl}}}\\
 &\leq & \underbrace{\Ex{}{|A|^3\cdot \indfun{|A|<\delta\sqrt{\bl}} } }_{\leq \Ex{}{|A|^3}}{}\frac{k_2}{6\bl^{3/2}}.\label{eq:app-bound-x-u-c3}
\end{IEEEeqnarray}
Here, (\ref{eq:app-bound-x-u-c3}) follows because $|\pdff''_B(x_0)|\leq k_2$.
Combining \eqref{eq:app-prob-x-leq-u} and \eqref{eq:taylor_expectation}, we obtain
\begin{IEEEeqnarray}{rCl}
\IEEEeqnarraymulticol{3}{l}{n^{3/2} \left|\prob[A\leq \sqrt{\bl}B] -\prob[B\geq 0] + \frac{\pdff_B'(0)}{2\bl} \right|}\notag\\
\;\; &=&n^{3/2} \Bigg| \int\nolimits_{|a|\geq\delta\sqrt{\bl}}\!\!\prob[B \geq a/\sqrt{\bl}] d\indist_A  \notag\\
&&\qquad\, +  \underbrace{(1-\cdfF_B(0))}_{\leq 1}\prob[|A|\geq \delta\sqrt{\bl}]\notag\\
&&\qquad \,-\, {f_B(0)\over \sqrt{\bl}}c_1(\bl) + {f_B'(0)\over 2\bl} c_2(\bl) - c_3(\bl)\Bigg|\label{eq:non-asy-bound-key-lemma1}\\
 &\leq&  \delta^{-3}\Ex{}{|A|^3} + \delta^{-3}\Ex{}{|A|^3} +  \frac{k_2\Ex{}{|A|^3}}{6}\notag\\
&& +\,k_2\underbrace{\big(\delta^{-2} + (2\delta)^{-1}\big) \Ex{}{ |A|^3 \cdot\indfun{|A|\geq \delta \sqrt{\bl}}}}_{\define c_4(\bl)}\IEEEeqnarraynumspace\label{eq:non-asy-bound-key-lemma2}\\
&= & (k_2/6 + 2\delta^{-3})\Ex{}{ |A|^3} +k_2 c_4(\bl)\label{eq:non-asy-bound-key-lemma}
\end{IEEEeqnarray}
where (\ref{eq:non-asy-bound-key-lemma2}) follows from  (\ref{eq:lemma_outer_term}), \eqref{eq:app-bound-x-u-c1} and (\ref{eq:app-bound-x-u-c2}).
The proof is concluded by taking $\bl\to\infty$ on each side of (\ref{eq:non-asy-bound-key-lemma}), and by using that
\begin{IEEEeqnarray}{rCl}
\lim\limits_{\bl\to\infty} c_4(\bl) =0.
\label{eq:non-asy-bound-key-lemma-vanishing-term}
\end{IEEEeqnarray}
}

\ifthenelse{\boolean{confversion}}
{
\section{}
\label{sec:proof-theorem}
To establish (\ref{eq:thm-R-star-expansion}), we study the converse bound (\ref{eq:thm-converse-rcsit}) and the achievability bound (\ref{eq:lb-numcode-simo_final}) in the large-$\bl$ limit.
Due to space limitations, we shall only provide a sketch of the proof of Theorem~\ref{thm:dispersion-simo}.
We refer the reader to~\cite{yang13-01a} for the missing steps.

Applying~\cite[Eq.~(102)]{polyanskiy10-05} to the RHS of (\ref{eq:converse-R}) yields
\begin{IEEEeqnarray}{rCl}
\Rcsirt^{\ast}(\bl-1,\error) \leq \frac{n}{n-1}\left(\gamma_\bl + \frac{\log \bl}{\bl}\right)
\end{IEEEeqnarray}
where $\gamma_\bl$ satisfies
\begin{IEEEeqnarray}{rCl}
\label{eq:proof-converse-asy-gamma}
\prob[ S_\bl \leq  \bl \gamma_{\bl}] = \error + 1/{\bl}.
\end{IEEEeqnarray}
To compute $\gamma_\bl$, note that---given $G$---the random variable $S_\bl$ is the sum of $\bl$ i.i.d. random variables with mean $\mu(G)\define \log(1+\snr G)$ and variance
$\sigma^2(G) \define\snr G(\snr G + 2)(1+\snr G)^{-2}$. An application of a Cramer-Esseen-type central-limit theorem \cite[Thm. VI.1]{petrov75} allows us to establish that~\cite{yang13-01a}
\begin{IEEEeqnarray}{rCl}
\prob[S_\bl\leq \bl\gamma_\bl] &=&\prob\lefto[Z \leq \sqrt{\bl} U(\gamma_\bl) \right] +\bigO\big(\bl^{-3/2}\big)\IEEEeqnarraynumspace\label{eq:app-prob-Sn-final-step1}
\end{IEEEeqnarray}
where $Z\sim \mathcal{N}(0,1)$ and $U(\gamma_\bl)\define(\gamma_\bl-\mu(G))/\sigma(G)$ are independent. Then, by Lemma~\ref{lem:expectation-phi},
\begin{IEEEeqnarray}{rCl}
\prob[S_\bl\leq \bl\gamma_\bl]&=& \underbrace{\prob[\mu(G)\leq \gamma_\bl]}_{=\cadist(\gamma_\bl)} + q(\gamma_\bl)/\bl + \bigO\big(\bl^{-3/2}\big)\IEEEeqnarraynumspace\label{eq:app-prob-Sn-final}
\end{IEEEeqnarray}
where $q(\gamma_\bl)\define\pdff_{U(\gamma_\bl)}'(0)$. Substituting (\ref{eq:app-prob-Sn-final}) into (\ref{eq:proof-converse-asy-gamma}), and applying Taylor's theorem to $\cadist(\gamma_\bl)$, we get
\begin{IEEEeqnarray}{rCl}
\gamma_\bl &=& C_{\error}+ \frac{\funcU(C_\error)+2}{2\bl} \cdot \frac{1}{F_C'(C_\error)} +\littleo(1/n)\label{eq:coh-final-step}.
\end{IEEEeqnarray}
Since $F'_C(C_\error)>0$ by assumption, we conclude that $\gamma_\bl=C_\error + \bigO(1/\bl)$. 

The analysis of the achievability bound follows similar steps~\cite{yang13-01a}.

}
{

\section{Proof of Theorem~\ref{thm:dispersion-simo}}
\label{sec:proof-theorem}
To establish Theorem~\ref{thm:dispersion-simo}, we study the converse bound (\ref{eq:thm-converse-rcsit}) and the achievability bound (\ref{eq:lb-numcode-simo_final}) in the large-$\bl$ limit.

\subsection{Converse} 
\label{sec:converse}
We begin by upper-bounding the RHS of~\eqref{eq:thm-converse-rcsit} by recalling that
for every $\gamma>0$~\cite[Eq.~(102)]{polyanskiy10-05}
 \begin{IEEEeqnarray}{c}\label{eq:inequality_beta}
 \alpha \leq \indist \lefto[\frac{d\indist}{d\outdist}\geq \gamma \right] + \gamma \beta_\alpha(\indist,\outdist).
 \end{IEEEeqnarray}
Using~\eqref{eq:inequality_beta} on~\eqref{eq:neyman_pearson_one} and setting $\gamma=e^{n\gamma_{\bl}}$ we obtain
\begin{IEEEeqnarray}{rCl}
\IEEEeqnarraymulticol{3}{l}{
 \beta_{1-\error}(\indist_0,\outdist_0)}\notag\\
 \,\,\,\, &\geq&  e^{-n\gamma_n}\big(1-\error - \indist_{0} \lefto[r(\pickcwd;\outseqrand\randvech) \geq n\gamma_n \right]\big)\IEEEeqnarraynumspace\\
 &=&   e^{-n\gamma_n}\big(\indist_{0} \lefto[r(\pickcwd;\outseqrand\randvech) \leq n\gamma_n \right]-\error\big).
 \IEEEeqnarraynumspace
 \end{IEEEeqnarray}
This allows us to upper-bound the RHS of~\eqref{eq:thm-converse-rcsit} as
\begin{IEEEeqnarray}{rCl}
\label{eq:thm-converse-rcsit-app}
\Rcsirt^{\ast}(\bl-1,\error) \leq \frac{n}{n-1}\left[\gamma_\bl - \frac{1}{\bl}\log\mathopen{}\big(\prob[ S_\bl \leq  \bl \gamma_{\bl}] - \error \big)\right]\IEEEeqnarraynumspace
\end{IEEEeqnarray}
for every~$\gamma_{\bl}$ satisfying
\begin{align}
   \prob[ S_\bl \leq  \bl \gamma_{\bl}] \geq \error.
\end{align}
We shall take $\gamma_\bl$ so that
\begin{IEEEeqnarray}{rCl}
\label{eq:proof-converse-asy-gamma}
\prob[ S_\bl \leq  \bl \gamma_{\bl}] = \error + \frac{1}{\bl}.
\end{IEEEeqnarray}
For this choice of $\gamma_\bl$, (\ref{eq:thm-converse-rcsit-app}) reduces to
\begin{IEEEeqnarray}{rCl}
\label{eq:conv-Rrt-gamma-n}
\Rcsirt^{\ast}(\bl-1,\error) \leq \frac{n}{n-1}\left[\gamma_\bl + \frac{\log \bl}{\bl}\right].
\end{IEEEeqnarray}
The proof is completed by showing that (\ref{eq:proof-converse-asy-gamma}) holds for
\begin{IEEEeqnarray}{rCl}
\label{eq:bound-gamma_n}
\gamma_\bl = C_\error + \bigO(1/\bl).
\end{IEEEeqnarray}
To prove (\ref{eq:bound-gamma_n}), we evaluate $\prob[ S_\bl \leq  \bl \gamma_{\bl}]$ in the limit $\bl\to\infty$ up to a~$\littleo(1/\bl)$ term.
Note that, given~$G$, the random variable~$S_\bl$ is the sum of~$\bl$ i.i.d. random variables with mean $\mu(G)\define \log(1+\snr G)$ and variance
\begin{IEEEeqnarray}{rCl}
\sigma^2(G) &\define& \frac{\snr G(\snr G + 2)}{(1+\snr G)^2}.
\end{IEEEeqnarray}
Hence,
\begin{IEEEeqnarray}{rCl}
\prob\lefto[S_\bl \leq \bl \argpn\right] &=& \prob\lefto[\frac{1}{\sqrt{\bl}}\sum\limits_{j=1}^{\bl} T_j \leq \sqrt{\bl}U(\argpn) \right]\label{eq:converse_info_densi_formula}
\end{IEEEeqnarray}
where
\begin{IEEEeqnarray}{rCl}
\label{eq:defn_tj}
T_j\define \frac{1}{\sigma(G)} \left(1-\frac{\big|\sqrt{\snr G}Z_i-1\big|^2}{1+\snr G}\right)
\end{IEEEeqnarray}
are zero-mean, unit-variance random variables that are conditionally independent given $G$, and\footnote{We shall write $U(\argpn)$ simply as $U$ whenever stressing its dependence on $\argpn$ is unnecessary.}
\begin{IEEEeqnarray}{c}
\label{eq:def_U}
U(\argpn) \define \frac{\argpn-\mu(G)}{\sigma(G)}.
\end{IEEEeqnarray}
 %
The following lemma, which is based on a Cramer-Esseen-type central-limit theorem~\cite[Thm.~VI.1]{petrov75} and on Lemma~\ref{lem:expectation-phi}, shows that~(\ref{eq:converse_info_densi_formula}) can be closely approximated by $\prob[U(\argpn)\geq 0]$.
\begin{lemma}
\label{lemma:prob_U}
 Let $\{T_j\}_{j=1}^{\bl}$ be given in (\ref{eq:defn_tj}) and let $U(\argpn)$ be given in (\ref{eq:def_U}) with $G$ satisfying the assumptions in Theorem~\ref{thm:dispersion-simo}.
 Take an arbitrary $\argpn_0>0$ that satisfies $\prob[U(\argpn_0) \geq 0] >0$.
 Then there exists a $\delta >0$ so that
\begin{IEEEeqnarray}{rCl}
\IEEEeqnarraymulticol{3}{l}{
\lim\limits_{\bl\to\infty} \sup\limits_{\argpn \in(\argpn_0-\delta, \argpn_0+ \delta)}  \bl^{3/2} \left|\prob\Bigg[\frac{1}{\sqrt{\bl}}\sum\limits_{j=1}^{\bl} T_j \leq \sqrt{\bl} U(\argpn)\Bigg]\right.} \notag\\
\quad &&\qquad\qquad\qquad\qquad\quad \left. \vphantom{\Bigg[\frac{1}{\sqrt{\bl}}\sum\limits_{j=1}^{\bl} \Bigg]}- \prob[U(\argpn)\geq 0] + \frac{\funcU(\argpn)}{2\bl}\right|< \infty\IEEEeqnarraynumspace\label{eq:lem-U-prob-expression}
\end{IEEEeqnarray}
where
\begin{IEEEeqnarray}{rCl}
\funcU(\argpn) &\define& \pdff'_{U(\argpn)}(0)\\
&=& -\frac{e^{2\argpn}\!-\!1}{\snr^2}\pdff'_G\lefto(\frac{e^\argpn\!-\!1}{\snr}\right) - \frac{e^{-\argpn}+e^{\argpn}}{\snr}\pdff_G\lefto(\frac{e^\argpn\!-\!1}{\snr}\right).\IEEEeqnarraynumspace\label{eq:def-funcU}
\end{IEEEeqnarray}
\end{lemma}
\begin{IEEEproof}
See Appendix~\ref{app:proof_U}.
\end{IEEEproof}

Note that
\begin{IEEEeqnarray}{rCl}
\prob[ U(\argpn)\geq 0 ] = \prob[\mu(G)\leq \argpn] = \cadist(\argpn)
\end{IEEEeqnarray}
where $\cadist(\argpn)$ is defined in (\ref{eq:def-cadist}). Hence, setting $\argpn = \gamma_\bl$ and $\argpn_0 = C_\error$, we get
\begin{IEEEeqnarray}{rCl}
\prob[S_\bl \leq \bl \gamma_\bl]
&=& \cadist(\gamma_\bl)  - \frac{\funcU(\gamma_\bl)}{2\bl} + \bigO(\bl^{-3/2})\label{eq:expression-pn}
\end{IEEEeqnarray}
where $\bigO(\bl^{-3/2})$ is uniform in $\gamma_\bl \in (C_\error-\delta, C_\error+\delta)$ for some $\delta>0$.
Substituting (\ref{eq:expression-pn}) into (\ref{eq:proof-converse-asy-gamma}), we finally obtain
\begin{IEEEeqnarray}{rCl}
\label{eq:equation_for_r_n}
\cadist(\gamma_\bl) - \frac{\funcU(\gamma_\bl)}{2\bl} + \bigO(\bl^{-3/2}) = \error + \frac{1}{\bl}.
\end{IEEEeqnarray}
By Taylor's theorem~\cite[Thm.~5.15]{rudin76}
\begin{equation}
\label{eq:expansion-FC-taylor}
\cadist(\gamma_\bl) = \cadist(C_\error) + \left(\left.\frac{d\cadist(\argpn)}{d\argpn}\right|_{\argpn = C_\error} + \littleo(1)\right)(\gamma_\bl-C_\error).
\end{equation}
Substituting (\ref{eq:expansion-FC-taylor}) into (\ref{eq:equation_for_r_n}) and using that $\cadist(C_\error)=\error$, we get
\begin{IEEEeqnarray}{c}
\label{eq:coh-final-step}
\gamma_\bl = C_{\error}+ \frac{\funcU(C_\error)+2}{2\bl} \cdot \frac{1}{\left.\frac{d\cadist(\argpn)}{d\argpn}\right|_{\argpn = C_\error} } +\littleo(1/n).
\end{IEEEeqnarray}
The proof of (\ref{eq:bound-gamma_n}) is concluded by noting that, by assumption, $\funcU(C_\error)<\infty$ and $\frac{d\cadist(\argpn)}{d\argpn}\Big|_{\argpn=C_\error} >0$.

\subsection{Achievability} 
\label{sec:achievability}
We set $\tau=1/\bl$ and $\gamma_\bl = \exp(-C_{\error} +\bigO(1/\bl))$ in~\eqref{eq:lb-numcode-simo_final} and we use that
\begin{IEEEeqnarray}{rCl}
F(\gamma_\bl;\bl-\rxant,\rxant)
&{\leq}& \frac{\Gamma(\bl)}{\Gamma(\bl-\rxant)\Gamma(\rxant)} \int\limits_{0}^{\gamma_\bl} t^{(\bl-\rxant)-1}dt\\
&=& \frac{\Gamma(\bl)}{\Gamma(\bl-\rxant + 1)\Gamma(r)} \gamma_\bl^{\bl-\rxant}\\
&\leq & \bl^{\rxant-1} \gamma_\bl^{\bl-\rxant}.
\end{IEEEeqnarray}
This yields,
\begin{IEEEeqnarray}{rCl}
\frac{\log \NumCode}{\bl} &\geq&  C_{\error} - \rxant\frac{\log(\bl)}{\bl} +\bigO\lefto(\frac{1}{n}\right).
\end{IEEEeqnarray}
To conclude the proof, we show that the choice $\gamma_\bl = \exp(-C_{\error} +\bigO(1/\bl))$  satisfies
\begin{IEEEeqnarray}{rCl}
 \indist_{\randmatY | \randvecx = \vecx_0}[\angletest_{\vecx_0}(\randmatY)=1] \geq 1-\error + 1/\bl.
\end{IEEEeqnarray}

Given $\randvech = \vech\neq \mathbf{0}$, we have that\footnote{Note that $\randvech =\mathbf{0}$ with zero probability.}
\begin{IEEEeqnarray}{rCl}
\cos\theta(\pickcwdnoch,\randmatY) &=& \max \limits_{\veca \in \complexset^{\rxant}\backslash\{\mathbf{0}\}} \frac{|\inprod{\pickcwdnoch}{\randmatY  \veca}|}{\|\pickcwdnoch\|\|\randmatY  \veca \|}\\
&\geq & \frac{|\inprod{\pickcwdnoch}{\randmatY\conjugate{\vech}}|}{\|\pickcwdnoch\|\|\randmatY  \conjugate{\vech} \|} .\label{eq:lower-bound-cos-theta}
\end{IEEEeqnarray}
Then
\begin{IEEEeqnarray}{rCl}
\IEEEeqnarraymulticol{3}{l}{
\indist_{\randmatY\given \randvecx = \pickcwdnoch}[\cos^2\theta(\pickcwdnoch, \randmatY)\geq 1-\gamma_\bl] }\notag\\
 \;\; &=& \Ex{\randvech}{
 \indist_{\randmatY\given \randvech=\vech, \randvecx =\pickcwdnoch}[\cos^2\theta(\pickcwdnoch, \randmatY)\geq 1-\gamma_\bl] }\\
 &\geq &\Ex{\randvech}{
 \indist_{\randmatY\given \randvech=\vech, \randvecx = \pickcwdnoch}\lefto[ \frac{|\inprod{\pickcwdnoch}{\randmatY\conjugate{\vech}}|^2}{\|\pickcwdnoch\|^2\|\randmatY  \conjugate{\vech} \|^2}\geq 1-\gamma_\bl\right] }\IEEEeqnarraynumspace\\
 &=&\indist_{\randmatY\randvech \given \randvecx = \pickcwdnoch}\lefto[\frac{|\inprod{\pickcwdnoch}{\randmatY\conjugate{\randvech}}|^2}{\|\pickcwdnoch\|^2\|\randmatY  \conjugate{\randvech} \|^2}\geq 1-\gamma_\bl \right].\label{eq:app-lower-bound-prob-cos-theta}
\end{IEEEeqnarray}
Under $\indist_{\randmatY\randvech \given \randvecx = \pickcwdnoch}$, the term $|\!\inprod{\pickcwdnoch}{\randmatY\conjugate{\randvech}}\!|^2/(\|\pickcwdnoch\|^2\|\randmatY \conjugate{\randvech} \|^2)$ is distributed as
\begin{IEEEeqnarray}{rCl}
\label{eq:app-distribution-cos-theta}
\frac{\left|\sqrt{\bl\snr}\|\randvech\|^2 + \bl^{-1/2}\sum\nolimits_{j=1}^{\bl}\herm{\randvecw_j}\randvech  \right|^2}{\sum\nolimits_{i=1}^{\bl}\big|\sqrt{\snr}\|\randvech\|^2+ \herm{\randvecw_j} \randvech \big|^2}
\end{IEEEeqnarray}
where $\randvecw_j\sim \jpg(\mathbf{0},\matI_\rxant)$.
Note that $\herm{\randvecw_j} \randvech$ has the same distribution as $\sqrt{G}Z_j$, where $Z_j\sim \jpg(0,1)$. Hence, the random ratio in (\ref{eq:app-distribution-cos-theta}) is distributed as
\begin{IEEEeqnarray}{rCl}
\frac{\left|\sqrt{\bl G\snr} +\bl^{-1/2}\sum\nolimits_{i=1}^{\bl}Z_i\right|^2}{\sum\nolimits_{i=1}^{\bl}\big|Z_i+\sqrt{G\snr}\big|^2}.
\end{IEEEeqnarray}
Therefore,
\begin{IEEEeqnarray}{rCl}
\IEEEeqnarraymulticol{3}{l}{
 \indist_{\randmatY\randvech \given \randvecx = \pickcwdnoch}\lefto[\frac{|\inprod{\pickcwdnoch}{\randmatY\conjugate{\randvech}}|^2}{\|\vecx\|^2\|\randmatY  \conjugate{\randvech} \|^2}\geq 1-\gamma_\bl \right] }\notag\\
 \;\; &=& \prob\lefto[\frac{\left|\sqrt{\bl G\snr} +\bl^{-1/2}\sum\nolimits_{i=1}^{\bl}Z_i\right|^2}{\sum\nolimits_{i=1}^{\bl}\big|Z_i+\sqrt{G\snr}\big|^2} \geq 1- \gamma_\bl \right]\\
 &=&  \prob\lefto[ \frac{\sum\nolimits_{i=1}^{\bl}|Z_i|^2 - \left|\sum_{i=1}^{\bl}Z_i\right|^2/\bl}{\sum_{i=1}^{\bl}\big|Z_i+\sqrt{G\snr}\big|^2 } \leq \gamma_\bl\right]\\
 &\geq &\prob\lefto[ \frac{\sum\nolimits_{i=1}^{\bl}|Z_i|^2}{\sum_{i=1}^{\bl}\big|Z_i+\sqrt{G\snr}\big|^2 } \leq \gamma_\bl\right]\\
 &{=}& \prob\lefto[ \sum\limits_{i=1}^{\bl}\left|(1-\gamma_\bl)Z_i - \gamma_\bl \sqrt{G\snr}\right|^2 \leq \bl \gamma_\bl G\snr\right]\\
 &=&   \prob\lefto[\frac{1}{\sqrt{\bl}}\sum\limits_{j=1}^{\bl}\altT_j \leq \sqrt{\bl} \altU\right]
 \label{eq:prob-noncoh-simo.}
\end{IEEEeqnarray}
where
\begin{IEEEeqnarray}{rCl}
\altU &\define& \frac{\gamma_\bl G\snr - \altmean(G)}{\sigma(G)} = \frac{\gamma_\bl(1+G\snr) - 1}{\sqrt{(1-\gamma_\bl)^2+2\gamma_{\bl}^2G\snr }}\IEEEeqnarraynumspace
\end{IEEEeqnarray}
and
\begin{IEEEeqnarray}{rCl}
\altT_j &\define&\frac{1}{\altvar(G)} \left(\left|(1-\gamma_\bl)Z_i - \gamma_\bl \sqrt{G\snr}\right|^2 - \altmean(G)\right)
\end{IEEEeqnarray}
with
\begin{IEEEeqnarray}{rCl}
\altmean(G) &\define& (1-\gamma_\bl)^2 +\gamma_{\bl}^2 G \snr
\end{IEEEeqnarray}
and
\begin{IEEEeqnarray}{rCl}
\altvar^2(G) &\define& (1-\gamma_\bl)^2\left[(1-\gamma_\bl)^2+2\gamma_{\bl}^2G\snr \right].
\end{IEEEeqnarray}
Note that, $\{\altT_j\}_{j=1}^{\bl}$ are zero-mean, unit-variance random variables that are conditionally independent given $G$.

To summarize, we showed that
\begin{IEEEeqnarray}{rCl}
\IEEEeqnarraymulticol{3}{l}{
 \indist_{\randmatY\given \randvecx = \vecx_0}[\cos^2\theta(\vecx_0, \randmatY)\geq 1-\gamma_\bl]}\notag\\
  \,\,\,&\geq& \prob\lefto[\frac{1}{\sqrt{\bl}}\sum\limits_{j=1}^{\bl}\altT_j \leq \sqrt{\bl} \altU\right].
\end{IEEEeqnarray}
To conclude the proof, it suffices to show that $\gamma_\bl=\exp(-C_{\error}+\bigO(1/\bl))$ yields
\begin{IEEEeqnarray}{rCl}
\label{eq:app-non-equal}
\prob\lefto[\frac{1}{\sqrt{\bl}}\sum\limits_{j=1}^{\bl}\altT_j \leq \sqrt{\bl} \altU\right]= 1-\error +\frac{1}{\bl}.
\end{IEEEeqnarray}
%
To this end, we proceed along the lines of the converse proof to obtain
\begin{equation}
\prob\lefto[{1\over\sqrt{\bl}}\sum\limits_{j=1}^{\bl}\altT_j \leq \sqrt{\bl} \altU\right] = \prob[\altU \geq 0]-\frac{\altf(\altgamma_{\bl})}{2\bl} + \bigO(\bl^{-3/2})
\label{eq:app-non-expansion-pn}
\end{equation}
where
\begin{IEEEeqnarray}{rCl}
\label{eq:app-noncoh-term-2}
\altf(\altgamma_{\bl})\define\pdff'_{\altU}(0) = \frac{(e^{\altgamma_{\bl}}-1)^2 }{\snr^2}\pdff'_G(\altg_0) + \frac{2}{\snr} \pdff_G(\altg_0)
\end{IEEEeqnarray}
with $\altgamma_\bl \define -\log\gamma_\bl$ and $\altg_0 \define(e^{\altgamma_{\bl}}-1)/\snr$, and where $\bigO(\bl^{-3/2})$ is uniform in $\tilde{\gamma}_\bl\in(C_\error-\delta, C_\error + \delta)$ for some $\delta>0$. We further have that
\begin{IEEEeqnarray}{rCl}
\label{eq:app-noncoh-term-1}
\prob[\altU \geq 0] = \prob[\log(1+G\snr) \geq \altgamma_\bl] = 1 - \cadist( \altgamma_\bl).
\end{IEEEeqnarray}
Substituting (\ref{eq:app-noncoh-term-2}) and (\ref{eq:app-noncoh-term-1}) into (\ref{eq:app-non-expansion-pn}), and then (\ref{eq:app-non-expansion-pn}) into (\ref{eq:app-non-equal}), we get
\begin{IEEEeqnarray}{rCl}
\cadist(\altgamma_\bl) + \frac{\altf\lefto(\altgamma_\bl\right)}{2\bl} +\bigO(\bl^{-3/2}) &=& \error - \frac{1}{\bl}.
\end{IEEEeqnarray}
Finally, using the same steps as in (\ref{eq:equation_for_r_n})--(\ref{eq:coh-final-step}), we obtain
\begin{IEEEeqnarray}{rCl}
\altgamma_\bl &=& C_{\error} - \frac{\altf \lefto(C_\error\right)+2}{2\bl}\frac{1}{\left.\frac{d\cadist(\argpn)}{d\argpn}\right|_{\argpn=C_{\error}}} + \littleo(1/\bl)\\
&{=}& C_\error + \bigO(1/\bl)\label{eq:ach-gamma-n-asy}
\end{IEEEeqnarray}
where (\ref{eq:ach-gamma-n-asy}) follows because $\altf\lefto(C_\error\right)<\infty$ and because $\left.\frac{d\cadist(\argpn)}{d\argpn}\right|_{\argpn=C_{\error}}>0$ by assumption.
This concludes the proof.

\section{Proof of Lemma~\ref{lemma:prob_U}}
\label{app:proof_U}
Fix $\argpn_0>0$ satisfying $\prob[U(\argpn_0)\geq 0]>0$.
Observe that
\begin{IEEEeqnarray}{rCl}
\prob[U(\argpn)\geq 0] = \prob[\log(1+\snr G)\leq \argpn] = \cadist(\argpn)
\end{IEEEeqnarray}
where $\cadist(\argpn)$ is defined in (\ref{eq:def-cadist}).
Since $\cadist(\argpn)$ is continuous in $\argpn$, there exists $0<\delta<\argpn_0$ so that
$\cadist(\argpn)>0$ (and hence, $\prob[U(\argpn)\geq 0] >0$) for every $\argpn \in (\argpn_0 - \delta, \argpn_0 + \delta)$.

To establish Lemma~\ref{lemma:prob_U}, we will need the following version of the Cramer-Esseen Theorem.\footnote{The Berry-Esseen Theorem used in \cite{polyanskiy10-05} to establish (\ref{eq:approx_R_introduction}) yields asymptotic expansions up to a $\bigO(1/\sqrt{\bl})$ term. This is not sufficient here, since we need to establish an asymptotic expansion up to a  $\littleo(1/\bl)$ term.}
%
%
\begin{thm}
\label{thm:refine-be}
Let $\{X_i\}_{i=1}^{\bl}$ be a sequence of i.i.d. real random variables having zero mean and unit variance.
Furthermore, let
\begin{IEEEeqnarray}{rCl}
v(t)&\define&\Ex{}{e^{itX_1}},\,\,\text{and}\,\,
F_\bl(x) \define \prob\lefto[\frac{1}{\sqrt{\bl}} \sum\limits_{j=1}^{\bl}X_j \leq x\right].\IEEEeqnarraynumspace
\end{IEEEeqnarray}
If $\Ex{}{|X_1|^4} < \infty$ and if $\sup_{|t|\geq \zeta} |v(t)| \leq \ConstThm$ for some $\ConstThm <1$, where $\zeta\define1/({12\Ex{}{|X_1|^3}})$, then for all $x$ and $\bl$
\begin{IEEEeqnarray}{rCl}
\IEEEeqnarraymulticol{3}{l}{\left|F_\bl(x) - Q(-x) - k_1(1-x^2)e^{-x^2/2}\frac{1}{\sqrt{\bl}}\right|}\notag\\
\quad\leq k_2\left\{n^{-1}(1+|x|)^{-4}\Ex{}{|X_1|^4} + \bl^6\left(k_0+\frac{1}{2\bl}\right)^{\bl}\right\}.\IEEEeqnarraynumspace
\label{eq:thm-osipov-refine}
\end{IEEEeqnarray}
Here, $k_1 \define \Ex{}{X_1^3}/(6\sqrt{2\pi})$, and $k_2$ is a positive constant independent of $\{X_i\}_{i=1}^{\bl}$ and $x$.

\end{thm}
\begin{IEEEproof}
The inequality~\eqref{eq:thm-osipov-refine} is a consequence of the tighter inequality reported in~\cite[Thm.~VI.1]{petrov75}.
\end{IEEEproof}
To prove Lemma~\ref{lemma:prob_U}, we proceed as follows. Note that
\begin{equation}
\label{eq:prob-equal-exp-conditional}
\prob\lefto[\frac{1}{\sqrt{\bl}}\sum\limits_{j=1}^{\bl} T_j \leq \sqrt{\bl}U\right] =\Ex{G}{\prob\lefto[\frac{1}{\sqrt{\bl}}\sum\limits_{j=1}^{\bl} T_j \leq \sqrt{\bl}U \Bigg| G\right]}.
\end{equation}
We next estimate the conditional probability on the RHS of (\ref{eq:prob-equal-exp-conditional}) using Theorem~\ref{thm:refine-be}.
 In order to do so, we need to verify that there exists a $k_0<1 $ such that $\sup_{g\in \NonnegReal}\sup_{|t|>\zeta} |v_{T_j}(t)|\leq k_0$, where $v_{T_j}(t)=\Ex{}{e^{itT_j}\given G=g}$. We start by evaluating $\zeta$. For all $g\in \NonnegReal$, it can be shown that
%
\begin{IEEEeqnarray}{rCl}
\label{eq:fourth_norm}
\Ex{}{|T_j|^4 \given G=g } = \frac{15(\snr g)^2+ 36\snr g + 12}{(\snr g+2)^2} \leq 15.
\end{IEEEeqnarray}
By Lyapunov's inequality~\cite[p.~18]{petrov75}, this implies that
\begin{IEEEeqnarray}{rCl}
\Ex{}{|T_j|^3 \given G=g} \leq \big(\Ex{}{|T_j|^4\given G=g}\big)^{3/4} \leq 15^{3/4}.\IEEEeqnarraynumspace
\end{IEEEeqnarray}
Hence,
\begin{equation}
\label{eq:app-zeta-bound}
\zeta = \frac{1}{12\Ex{}{|T_j|^3\given G=g}}\geq\frac{15^{-3/4}}{12}\define \zeta_0.
\end{equation}
By~\eqref{eq:app-zeta-bound}, we have that
\begin{IEEEeqnarray}{rCl}
\sup\limits_{|t|>\zeta} |v_{T_j}(t)| \leq \sup\limits_{|t|>\zeta_0} |v_{T_j}(t)|
\end{IEEEeqnarray}
where $\zeta_0$ does not depend on $g$.
We now compute $|v_{T_j}(t)|$. Observe that given $G=g$
\begin{equation}
T_j = \frac{1}{\sigma(g)} - \frac{\snr g}{2\sigma(g)(1+\snr g)} \underbrace{\left|\sqrt{2} Z_j -\sqrt{\frac{2}{\snr g}}  \right|^2}_{\define N_j}
\end{equation}
where the term $N_j$ follows a noncentral $\chi^2$ distribution with two degrees of freedom and noncentrality parameter $2/(\snr g)$.
If we let $v_{N_j}\define\Ex{}{e^{i t N_j}}$, then
\begin{IEEEeqnarray}{rCl}
|v_{T_j}(t)|&=& \left|\exp\lefto(\frac{i t}{\sigma(g)}\right)\right| \cdot \left|\Ex{}{\exp\lefto(  \frac{-it \snr  g  N_j}{2\sigma(g)(1+\snr g)}\right)}\right|\\
&=&  \left|v_{N_j}\lefto( \frac{-\snr g t}{2\sigma(g)(1+\snr g)}\right)\right|\\
&=&\exp\lefto(-\frac{t^2}{\snr gt^2 + \snr g+2}\right) \left(1+\frac{\snr g t^2}{\snr g + 2}\right)^{-1/2}\IEEEeqnarraynumspace\label{eq:compute-v-Tj}
 \end{IEEEeqnarray}
where (\ref{eq:compute-v-Tj}) follows from \cite[p.~24]{muirhead05}.
Now, observe that the RHS of (\ref{eq:compute-v-Tj}) is monotonically decreasing in $t$ and monotonically increasing in $g$.
Hence,
 \begin{IEEEeqnarray}{rCl}
 \IEEEeqnarraymulticol{3}{l}{\sup\limits_{g\in\NonnegReal }\sup\limits_{|t|\geq \zeta_0} |v_{T_j}(t)|}\notag\\
 &=&\! \sup\limits_{g\in\NonnegReal }\sup\limits_{|t|\geq \zeta_0}\!\!\left\{\exp\lefto(\!-\frac{t^2}{\snr g t^2\! + \!\snr g \!+\! 2}\!\right)\!\left(1\!+\!\frac{\snr g t^2}{\snr g \!+ \!2}\right)^{-1/2}\!\right\} \IEEEeqnarraynumspace\\
 &=&\!\sup\limits_{g\in\NonnegReal } \!\left\{\exp\lefto(\!-\frac{\zeta_0^2}{\snr g \zeta_0^2 + \snr g + 2} \right)\! \left(1+\frac{\snr g \zeta_0^2}{\snr g + 2}\right)^{-1/2}\right\}\\
 &\leq & \frac{1}{\sqrt{1+\zeta_0^2}}<1. \label{eq:app-upper-bound-cf}
 \end{IEEEeqnarray}
Set $\ConstThm = {1}/{\sqrt{1+\zeta_0^2}}$.
As we verified that the conditions in Theorem~\ref{thm:refine-be} are met, we conclude that for all $\bl$
\begin{IEEEeqnarray}{rCl}
\IEEEeqnarraymulticol{3}{l}{
\left|\prob\lefto[\frac{1}{\sqrt{\bl}}\sum\limits_{j=1}^{\bl} T_j \leq \sqrt{\bl}U\right] -\Ex{}{Q\lefto(-\sqrt{\bl}U\right)}\right|}\notag\\
\quad&\leq& \frac{k_3}{\sqrt{\bl}} \left|\Ex{}{(1-\bl U^2)e^{-\bl U^2/2}}\right| \notag\\
&&+\, \frac{k_4}{\bl} \Ex{}{(1+|\sqrt{\bl}U|)^{-4}}+  k_2 \bl^6\left(k_0+\frac{1}{2\bl}\right)^{\bl}\IEEEeqnarraynumspace\label{eq:app-estimate-prob-minus-phi}
\end{IEEEeqnarray}
where $k_3 \define 15^{3/4}/(6\sqrt{2\pi})$
and $k_4 \define 15 k_2$.
In view of~\eqref{eq:lem-U-prob-expression}, we note that
the last term on the RHS of (\ref{eq:app-estimate-prob-minus-phi}) satisfies
\begin{IEEEeqnarray}{rCl}
\label{eq:lemma-last-term-cramer-esseen}
\lim\limits_{\bl\to\infty} \bl^{3/2} \left( k_2 \bl^6\left(k_0+\frac{1}{2\bl}\right)^{\bl}\right) = 0.
\end{IEEEeqnarray}

Next, we prove the following two estimates
\begin{IEEEeqnarray}{rCl}
&&\lim\limits_{\bl\to\infty }\sup\limits_{\argpn\in(\argpn_0-\delta, \argpn_0 +\delta)} \sqrt{\bl} \Ex{}{\big(1+|\sqrt{\bl}U(\argpn)|\big)^{-4}} \leq k_5\IEEEeqnarraynumspace \label{eq:app-second-estimate-exp1}
\end{IEEEeqnarray}
\begin{IEEEeqnarray}{rCl}
&&\lim\limits_{\bl\to\infty }\sup\limits_{\argpn\in(\argpn_0-\delta, \argpn_0 +\delta)} \!\!\bl \left|\Ex{}{\!\big(1\!-\!\bl (U(\argpn))^2\big)e^{-\frac{\bl (U(\argpn))^2}{2}}}\!\right| \leq k_6\IEEEeqnarraynumspace \label{eq:app-second-estimate-exp}
\end{IEEEeqnarray}
for some constants $k_5,k_6<\infty$.
Note that since the map
\begin{align}
  (g,\argpn) \mapsto \left(\frac{\argpn-\mu(g)}{\sigma(g)},\argpn\right)
\end{align}
is a diffeomorphism (of class~$C^3$)~\cite[p.~147]{munkres91-a} in the region $\argpn>0,g>0$, the pdf $\pdff_{U(\argpn)}(t)$ of $U(\argpn)$ and its first and second derivative are jointly continuous functions of $(\argpn,t)$, and, hence, bounded on bounded sets.
Specifically,
for every $\argpn\in(\argpn_0 -\delta,\argpn_0 +\delta)$ and every $\altdelta>0$ there exists a $\altk <\infty$ so that
\begin{IEEEeqnarray}{rCl}
&&\sup_{t\in [-\altdelta, \altdelta]}\sup\limits_{\argpn\in(\argpn_0-\delta,\argpn_0+\delta)} |\pdff_{U(\argpn)}(t)| \leq \altk\label{eq:lemma-ub-pdfu} \\
&&\sup_{t\in [-\altdelta, \altdelta]}\sup\limits_{\argpn\in(\argpn_0-\delta,\argpn_0+\delta)} |\pdff_{U(\argpn)}'(t)| \leq \altk \label{eq:lemma-ub-pdfud}\\
&&\sup_{t\in [-\altdelta, \altdelta]}\sup\limits_{\argpn\in(\argpn_0-\delta,\argpn_0+\delta)} |\pdff_{U(\argpn)}''(t)| \leq \altk.\label{eq:lemma-ub-pdfudd}
\end{IEEEeqnarray}

%

Fix now $\altdelta>0$ and let $\altk$ as in~\eqref{eq:lemma-ub-pdfu}--\eqref{eq:lemma-ub-pdfudd}.
To prove~(\ref{eq:app-second-estimate-exp1}), we proceed as follows:
\begin{IEEEeqnarray}{rCl}
\IEEEeqnarraymulticol{3}{l}{
\Ex{}{\big(1+|\sqrt{\bl}U|\big)^{-4}}}\notag\\
\quad &=& \Ex{}{(1+|\sqrt{\bl}U|)^{-4} \indfun{|U|<\altdelta}}\notag\\
&&+\;\Ex{}{\big(1+|\sqrt{\bl}U|\big)^{-4} \indfun{|U|\geq \altdelta}}\\
&\leq &  2 \altk \int\nolimits_{0}^{\altdelta}(1 + \sqrt{\bl} t )^{-4} dt + (1+\sqrt{\bl}{\altdelta})^{-4}\label{eq:proof-lemma-estimate-mid}\\
& = & \frac{2\altk}{3\sqrt{\bl}} \left(1-(1+\sqrt{\bl}\altdelta)^{-3} \right)+ (1+\sqrt{\bl}{\altdelta})^{-4}\\
& \leq &  \frac{2\altk}{3\sqrt{\bl}} + \frac{1}{\bl^2\altdelta^4}
\end{IEEEeqnarray}
where in (\ref{eq:proof-lemma-estimate-mid}) we used (\ref{eq:lemma-ub-pdfu}). This proves \eqref{eq:app-second-estimate-exp1}.
The inequality (\ref{eq:app-second-estimate-exp}) can be established as follows. First, for $\bl \geq  \altdelta^{-2}$,
%
%
\begin{IEEEeqnarray}{rCl}
\IEEEeqnarraymulticol{3}{l}{
\left|\Ex{}{(1-\bl U^2)e^{-\bl U^2/2}}\right|} \notag\\
\;\;&\leq& \underbrace{\left|\int\nolimits_{-\altdelta}^{\altdelta}(1-\bl t^2)e^{-\bl t^2/2} \pdff_U(t) dt\right|}_{\define I_1}\notag\\
&&+ \,\underbrace{\Ex{}{(\bl U^2-1)e^{-\bl U^2/2} \indfun{|U|\geq \altdelta}}}_{\define I_2}.\label{eq:app-bound-exp-term-2}
\end{IEEEeqnarray}
To evaluate $I_1$, we use the relation
 $(1-\bl t^2)e^{-nt^2/2} = \frac{d}{dt} \lefto(t e^{-nt^2/2}\right)$ and integration by parts to obtain
\begin{IEEEeqnarray}{rCl}
I_1 &=& \left| \left.\left(t e^{-\bl t ^2/2} \pdff_U(t)\right)\right|_{-\altdelta}^{\altdelta} - \int\nolimits_{-\altdelta}^{\altdelta}t e^{-\bl t^2/2} \pdff'_U(t) dt  \right|\IEEEeqnarraynumspace\\
&\leq& 2 \altk \altdelta e^{-\bl \altdelta^2/2} + 2 \altk \frac{1}{\bl}\big(1 - e^{-\bl\altdelta^2/2}\big).
\end{IEEEeqnarray}
Therefore,
\begin{IEEEeqnarray}{rCl}
\label{eq:app-bound-I1}
\lim\limits_{\bl\to\infty } \sup\limits_{\argpn\in(\argpn_0-\delta, \argpn_0 +\delta)} \bl I_1 \leq 2\altk.
\end{IEEEeqnarray}
For $I_2$ we proceed as follows:
\begin{IEEEeqnarray}{rCl}
I_2 &\leq & \Ex{}{\bl U^2 e^{-\bl U^2/2} \cdot \indfun{|U|\geq \altdelta} }\\
&\leq & \sup\limits_{|t|\geq \altdelta} \big\{\bl t^2 e^{-\bl t^2/2} \big\}.
\end{IEEEeqnarray}
Note that when $ \bl > 2\altdelta^{-2} $, the function $\bl t^2 e^{-\bl t^2/2}$ is monotonically decreasing in $t\in[\altdelta,+\infty)$. Hence,
\begin{IEEEeqnarray}{rCl}
\label{eq:app-bound-I2}
\lim\limits_{\bl\to\infty } \sup\limits_{\argpn\in(\argpn_0-\delta, \argpn_0 +\delta)} \bl I_2 \leq \lim\limits_{\bl\to\infty } \bl^2\altdelta^2e^{-\bl\altdelta^2/2 } = 0.
\end{IEEEeqnarray}
Substituting (\ref{eq:app-bound-I1}) and (\ref{eq:app-bound-I2}) into (\ref{eq:app-bound-exp-term-2}), we obtain (\ref{eq:app-second-estimate-exp}).

%
Combining (\ref{eq:app-second-estimate-exp1}) and (\ref{eq:app-second-estimate-exp}) with \eqref{eq:app-estimate-prob-minus-phi}, we conclude that
\begin{IEEEeqnarray}{rCl}
\IEEEeqnarraymulticol{3}{l}{\lim \limits_{ \bl\to\infty } \sup\limits_{\argpn\in(\argpn_0-\delta, \argpn_0 +\delta)} n^{3/2} \left|\prob\lefto[\frac{1}{\sqrt{\bl}}\sum\limits_{j=1}^{\bl} T_j \leq \sqrt{\bl} U(\argpn)\right]\right.}\notag\\\left. \quad\quad\quad\quad\quad\quad\quad\quad\quad - \Ex{}{Q(-\sqrt{\bl}U(\argpn))} \vphantom{\prob\lefto[\frac{1}{\sqrt{\bl}}\sum\limits_{j=1}^{\bl} T_j \leq \sqrt{\bl} U(\argpn)\right]}\right| & \leq& k_5+k_6.\IEEEeqnarraynumspace
\end{IEEEeqnarray}
To conclude the proof of Lemma~\ref{lemma:prob_U}, we need to show that there exists a constant $k_7<\infty$ such that
\begin{IEEEeqnarray}{rCl}
 \IEEEeqnarraymulticol{3}{l}{\lim \limits_{ \bl\to\infty } \sup\limits_{\argpn\in(\argpn_0-\delta, \argpn_0 +\delta)} \!\!n^{3/2} \left|\vphantom{ \frac{\pdff'_{U(\argpn)}(0)}{2\bl}} \Ex{}{Q(-\sqrt{\bl}U(\argpn))} \right.}\notag\\ \left.\quad\quad\quad\quad\quad\quad\quad \quad\quad- \prob[U(\argpn)\geq 0]  +  \frac{\pdff'_{U(\argpn)}(0)}{2\bl}  \right|&& \leq k_7\IEEEeqnarraynumspace
\end{IEEEeqnarray}
where $\pdff_{U(\argpn)}$ is the pdf of $U(\argpn)$.
This follows by the uniform bounds~\eqref{eq:lemma-ub-pdfu}--\eqref{eq:lemma-ub-pdfudd}, and by (\ref{eq:non-asy-bound-key-lemma}). Note, in fact that the term $c_4(\bl)$ in the proof of Lemma~\ref{lem:expectation-phi}, when evaluated for \mbox{$Y=U(\argpn)$}, does not depend on $\argpn$.

Since $U(\argpn) = (\argpn-\mu(G))/\sigma(G)$, we get after algebraic manipulations
\begin{IEEEeqnarray}{rCl}
\funcU(\argpn) &=& \pdff'_{U(\argpn)}(0)\\
&=& -\frac{e^{2\argpn}\!-\!1}{\snr^2}\pdff'_G\lefto(\frac{e^\argpn\!-\!1}{\snr}\right) - \frac{e^{-\argpn}+e^{\argpn}}{\snr}\pdff_G\lefto(\frac{e^\argpn\!-\!1}{\snr}\right).\IEEEeqnarraynumspace
\end{IEEEeqnarray}
This concludes the proof.
}

\bibliographystyle{IEEEtran}
\bibliography{IEEEabrv,publishers,confs-jrnls,WeiBib}

\end{document}